\documentclass[11pt]{article}

\usepackage{amsmath}
\usepackage{graphicx}
\usepackage{amsfonts}
\usepackage{amssymb}
\usepackage{epsfig}
\usepackage{color}
\usepackage{psfrag}

\newcommand{\EQ}{\begin{equation}}
\newcommand{\EN}{\end{equation}}
\newcommand{\be}{\begin{equation}}
\newcommand{\ee}{\end{equation}}
\newcommand{\bea}{\begin{eqnarray}}
\newcommand{\eea}{\end{eqnarray}}

\begin{document}

\setcounter{page}{0} \topmargin0pt \oddsidemargin5mm \renewcommand{%
\thefootnote}{\arabic{footnote}}\newpage \setcounter{page}{0} 
\begin{titlepage}
\begin{flushright}
%OUTP-96-19S\\
SISSA 19/2011/EP
\end{flushright}
\vspace{0.5cm}
\begin{center}
{\large{\bf Potts $q$-color field theory and scaling random cluster model}
%\\{\bf in two-dimensional quantum field theory}
}\\

\vspace{1.8cm}
{\large Gesualdo Delfino
%$^{a,b}$ 
and Jacopo Viti
%$^{c\,\,*}$
} 
\\
\vspace{0.5cm}
{\em SISSA - Via Bonomea 265 - 34136, Trieste, Italy}\\
{\em INFN sezione di Trieste, Italy}\\
%{\em ${}^c$LPTM, Universit\'e de Cergy-Pontoise }\\
%{\em 2 avenue Adolphe Chauvin, 95302 Cergy-Pontoise, France}\\
%{\em E-mail: delfino@sissa.it, giuliano.niccoli@desy.de}\\
\end{center}
\vspace{1.2cm}

\renewcommand{\thefootnote}{\arabic{footnote}}
\setcounter{footnote}{0}

\begin{abstract}
\noindent
We study structural properties of the $q$-color Potts field theory which, for real values of $q$, describes the scaling limit of the random cluster model. We show that the number of independent $n$-point Potts spin correlators coincides with that of independent $n$-point cluster connectivities and is given by generalized Bell numbers. Only a subset of these spin correlators enters the determination of the Potts magnetic properties for $q$ integer. The structure of the operator product expansion of the spin fields for generic $q$ is also identified. For the two-dimensional case, we analyze the duality relation between spin and kink field correlators, both for the bulk and boundary cases, obtaining in particular a sum rule for the kink-kink elastic scattering amplitudes.
\end{abstract}

%\vspace{8cm}
%$^*$\,Present address: DESY Theory, Notkestr. 85, 22607 Hamburg, Germany.
\end{titlepage}

\newpage
\section{Introduction}
There are cases in which the notion of spontaneous symmetry breaking which is commonly used to describe ferromagnetic phase transitions admits an extension which allows the description of phase transitions of a different nature. The case of interest in this paper is that of the $q$-state Potts model \cite{Potts,Wu}, whose Hamiltonian
\EQ
{\cal H}=-J\sum_{\langle x,y\rangle}\delta_{s(x),s(y)}\,,\hspace{1.5cm}s(x)=1,\ldots,q\,,
\label{H}
\EN
is invariant under the group $S_q$ of permutations of the values of the variables $s(x)$ located at each site of a lattice ${\cal L}$. The $q$ values of $s(x)$ can conveniently be thought as $q$ different colors and, clearly, the Potts ferromagnet is well defined for an integer number of colors. It has been known for a long time, however, that the partition function $\sum_{\{s(x)\}}e^{-{\cal H}}$ can be rewritten, up to an inessential constant, as \cite{KF}
\EQ
Z=\sum_{\mathcal{G}\subseteq{\cal L}} p^{n_b}(1-p)^{\bar{n}_b}q^{N_c}\,,
\label{FK_rep}
\EN 
where $\mathcal{G}$ is a graph obtained
putting $n_b$ bonds on the lattice $\mathcal{L}$, each one with
weight $p=1-e^{-J}\in[0,1]$ ($\bar{n}_b$ is the number
of absent bonds in $\mathcal{L}$). The connected components of $\mathcal{G}$, including isolated sites, are called Fortuin-Kasteleyn (FK) clusters. In the formulation (\ref{FK_rep}), which defines the so called random cluster model, the probability measure for the graph $\mathcal{G}$ depends on $q$ through the factor $q^{N_c}$, $N_c$ being the number of clusters in $\mathcal{G}$, and is well defined for any real positive $q$. In the thermodynamic limit the random cluster model undergoes, for $q$ continuous, a percolative phase transition associated to the appearance, for $p$ larger than a $q$-dependent critical value $p_c$, of a non-zero probability of finding an infinite cluster. The transition is first order for $q$ larger than a dimensionality-dependent value $q_c$, and second order otherwise; in particular, the limit $q\to 1$, which eliminates the  factor $q^{N_c}$ in (\ref{FK_rep}), describes ordinary percolation. When $q$ is an integer larger than 1, the percolative transition of the random cluster model contains, in a sense to be clarified below, the ferromagnetic transition of the Potts model. 

For $q\leq q_c$, when a scaling limit exists, the problem admits a field theoretical formulation. There must be a field theory, which we call Potts field theory, that describes the scaling limit of the random cluster model for $q$ real, as well as that of the Potts ferromagnet for $q$ integer. This theory has the Potts spin fields as fundamental fields (the FK mapping relates Potts spin correlators and connectivities for the FK clusters) and is characterized by $S_q$-invariance. The obvious question of the meaning of $S_q$ symmetry for $q$ non-integer arose at least since the $\epsilon$ expansion treatment of \cite{PL}, and appears to admit a general answer: although one starts from expressions which are formally defined only for $q$ integer, formal use of the symmetry unambiguously leads to final expressions containing $q$ as a parameter which can be taken continuous.

The two-dimensional case allows for the most advanced, non-perturbative results. The Potts field theory is integrable and the underlying scattering theory was exactly solved in \cite{CZ} for continuous $q\leq q_c=4$. The scattering solution was then used in \cite{DC,DVC} to determine two-point correlators and universal combinations of critical amplitudes for the Potts model and percolation. In spite of these results, even in two dimensions the nature of Potts field theory for real values of $q$ needs to be better understood. Indeed, already in the critical case, although exact formulae for the scaling dimensions and the central charge as functions of $q$ have been known for a long time \cite{Nienhuis,DF}, and results for boundary correlators have been obtained since the work of \cite{Cardy}, a complete conformal field theory description is only available for the magnetic properties at $q=2,3,4$. The essential problem is readily stated: since the Potts spin field has $q-1$ independent components, no ordinary characterization of the space of fields appears possible for $q$ non-integer. Recently \cite{DV2}, we showed how the problem of non-integer field multiplicities can be dealt with for three-point functions, and arrived in this way to exact predictions for operator product expansion (OPE) coefficients which have already been confirmed by high-precision Monte Carlo simulations for the case of ordinary percolation \cite{ZSK,SZK}. 

In this paper we investigate, for $q$ continuous, some structural properties of the Potts field theory as a theory characterized by $S_q$ invariance under color permutations and able to describe the scaling limit of the random cluster model. We first of all observe that the issue of the content of the theory is better addressed, in any dimension, focusing on linearly independent correlation functions rather than on field multiplicities. For this purpose one needs to have in mind the relation of spin correlators with cluster connectivities for $q$ real, rather than with magnetic properties for $q$ integer. Just to make an example already considered in \cite{DV2}, the correlator of three spin fields with the same color is proportional to the probability that three points are in the same FK cluster. This probability is well defined and non-vanishing for continuous values of $q$ in the random cluster model, in particular for the case $q=2$ in which the spin correlator has a zero enforcing Ising spin reversal symmetry; stripped of a trivial factor $q-2$, this spin correlator enters the description of cluster connectivity at $q=2$ as for generic real values of $q$. Similarly, a number $n_c\leq n$ of different colors enters the generic $n$-point spin correlator: a correlator with $n_c>q$ has no role in the description of the Potts ferromagnet, but enters the determination of cluster connectivities in the random cluster model. One then realizes that the dimensionality $F_n$ of the space of linearly independent $n$-point spin correlators for real values of $q$ is actually $q$-independent and must coincide with that of the space of linearly independent $n$-point connectivities. We show that this is indeed the case and that $F_n$ coincides with the number\footnote{We consider the symmetric phase, i.e. the case $p\leq p_c$.} of partitions of a set of $n$ elements into subsets each containing more than one element; the relation between spin correlators and cluster connectivities is also given and written down explicitly up to $n=4$. Only a number $M_n(q)$, smaller than $F_n$ for $n$ large enough, of independent spin correlators enters the determination of the magnetic properties at $q$ integer, making clear that the magnetic theory is embedded into the larger percolative theory\footnote{See \cite{isingperc,DV1} for detailed studies of this fact in the Ising model.}.

Once the relevant correlation functions have been identified, an essential tool for their study is the OPE of the spin fields. Again, the existence of such an object for real values of $q$ is made a priori not obvious by the badly defined multiplicity of the fields. We show, however, that its structure can be very naturally identified (equations (\ref{OPE_sigma_ab}) and (\ref{OPE_sigma_aa}) below).

The additional property we study in this paper, duality, is specific of the two-dimensional case. It is well known \cite{Potts,Wu} that spin correlators computed in the symmetric phase of the square lattice Potts model coincide with disorder correlators computed in the spontaneously broken phase. Here we study duality directly in the continuum, for real values of $q$, and with the main purpose of clarifying the role of kink fields. These are the fields that in any two-dimensional field theory with a discrete internal symmetry create the kink excitations interpolating between two degenerate vacua of the spontaneously broken phase; in general, they are linearly related to the usual disorder fields, which are mirror images of the spin fields, in the sense that they carry the same representation of the symmetry. In the Potts field theory, however, it appears that in many respects the use of kink fields provides a simpler way of dealing with the symmetry for real values of $q$. The duality between spin and kink field $n$-point correlators is a non-trivial problem that we study in detail up to $n=4$, both for bulk and boundary correlations. The problem of correlation functions for points located on the boundary of a simply connected domain is simplified by topological constraints which, in the limit in which the boundary is moved to infinity, also account for non-trivial relations among kink scattering amplitudes.

The paper is organized as follows. In the next section we investigate cluster connectivities and Potts spin correlators, their multiplicity and the relation between them. In section~3 we analyze OPE's and obtain in particular that for the Potts spin fields for real values of $q$. Duality between spin and kink field correlators in two dimensions is studied in general in section~4 and specialized to boundary correlations in section~5. Few final remarks are given in section~6 and three appendices complete the paper.

\section{Counting correlation functions}
\subsection{Cluster connectivities}
Correlations within the random cluster model (\ref{FK_rep}) are expressed by the connectivity functions giving the probability that $n$ points $x_1,\ldots,x_n$ fall into a given FK cluster configuration. In order to define the connectivities we associate to a point $x_i$ a label $a_i$, with the convention that two points $x_i$ and $x_j$ belong to the same cluster if $a_i=a_j$, and to different clusters otherwise. We then use the notation $P_{a_1...a_n}(x_1,..., x_n)$ for the generic $n$-point connectivity function, within the phase in which there is no infinite cluster, i.e. for $p\leq p_c$. The total number of functions $P_{a_1\dots a_n}(x_1,\dots,x_n)$ is the number $B_n$ of possible partitions (clusterizations) of the $n$ points\footnote{The $B_n$'s are known as Bell numbers and are discussed in Appendix~A.}; these $B_n$ functions sum to one and form a set that we call $\mathcal{C}^{(n)}$. 

It is not difficult to realize that the elements of $\mathcal{C}^{(n)}$ can be rewritten as linear combinations of ``basic'' $k$-point connectivities, with $k=2,\ldots,n$; we call $F_k$ the number of basic $k$-point connectivities, and $\mathcal{P}^{(k)}\subset\mathcal{C}^{(k)}$ the set they form. There is a simple procedure to build $\mathcal{P}^{(n)}$ given the $\mathcal{P}^{(k)}$'s for $k=2,\dots,n-1$. Let us start with $n=2$:
$\mathcal{C}^{(2)}=\{P_{aa}, P_{ab}\}$, but $P_{aa}+P_{ab}=1$ implies $F_2=1$ and we can choose $\mathcal{P}^{(2)}=\{P_{aa}\}$. Consider now  $n=3$:
$\mathcal{C}^{(3)}=\{P_{aaa}, P_{aab},
P_{aba}, P_{baa}, P_{abc}\}$, however starting from $P_{aaa}$ and summing
over the inequivalent configurations of the last point we have $P_{aaa}+P_{aab}=P_{aa0}$, where $P_{aa0}\equiv P_{aa}(x_1,x_2)$ belongs, we chose it on purpose, to $\mathcal{P}^{(2)}$. We observe that only one linear
combination of connectivities in $\mathcal{C}^{(3)}$ reproduces, taking into account the coordinate dependence, the two-point connectivity in $\mathcal{P}^{(2)}$; we obtain then the sum rules
\begin{align}
&P_{aaa}+P_{aab}=P_{aa0}\equiv P_{aa}(x_1,x_2),\label{red_1}\\
&P_{aaa}+P_{aba}=P_{a0a}\equiv P_{aa}(x_1,x_3),\label{red_2}\\
&P_{aaa}+P_{baa}=P_{0aa}\equiv P_{aa}(x_2,x_3),\label{red_3}\\
&P_{aaa}+P_{aab}+P_{aba}+P_{baa}+P_{abc}=1.\label{red_4}
\end{align}
This system of equations exhausts all the possible linear relations among the elements of $\mathcal{C}^{(3)}$; a reduction to a non-basic two-point connectivity (i.e. not belonging to ${\cal P}^{(2)}$, for example $P_{ab0}$) will indeed produce an equation which is a linear combination of those above. It follows, in particular, that the five elements of $\mathcal{C}^{(3)}$ can be written in terms of $P_{aaa}, P_{aa0}, P_{0aa}, P_{a0a}$, so that $F_3=1$ and  we can choose $\mathcal{P}^{(3)}=\{P_{aaa}\}$.

In general, suppose we have chosen the $F_k$ basic connectivities in $\mathcal{P}^{(k)}$ for $k=2,\dots, n-1$. Given $P_{a_1\dots a_n}\in\mathcal{C}^{(n)}$, we can fix $k$ of its $n$ indices according to an
element of $\mathcal{P}^{(k)}$;  summing over the inequivalent configurations of the remaining $n-k$ indices we  obtain a linear relation for the connectivities of $\mathcal{C}^{(n)}$. We can do this for each of the $F_k$ elements in ${\cal P}^{(k)}$ and for $\binom{n}{k}$ choices of $k$ indices among $n$ indices.  The number of independent sum rules for the elements of $\mathcal{C}^{(n)}$ is then
\EQ
E_n=\binom{n}{n-1}F_{n-1}+\binom{n}{n-2}F_{n-2}+...+\binom{n}{2}F_2+1\,,
\label{ind_sum_rule}
\EN
with the last term accounting for the fact that the $n$-point connectivities sum to one. The number $F_n$ of elements of $\mathcal{P}^{(n)}$ is the minimum number of connectivities in $\mathcal{C}^{(n)}$ needed to solve the linear system of $E_n$ equations in $B_n$ unknowns, i.e.
\EQ
F_n=B_n-\binom{n}{n-1}F_{n-1}-\binom{n}{n-2}F_{n-2}-...-\binom{n}{2}F_2-1\,.
\label{ind_prob}
\EN  
Defining $F_0\equiv 1$ and knowing that $F_1=0$, we rewrite (\ref{ind_prob}) as  
\EQ
B_{n}=\sum_{k=0}^n\binom{n}{k}F_k\,,\quad\forall n\geq 0.
\label{atleast2_part}
\EN  
We show in Appendix~A that (\ref{atleast2_part}) implies that $F_n$ is the number of partitions of a set of $n$ points into subsets containing at least two points. We list in Table~\ref{tab_Bell} the first few $B_n$ and $F_{n}$. The combinatorial interpretation of the $F_n$'s suggests that a natural choice for the set $\mathcal{P}^{(n)}$ of linearly independent $n$-point connectivities is to consider clusterizations with no isolated points, i.e. $\mathcal{P}^{(2)}=\{P_{aa}\}$, $\mathcal{P}^{(3)}=\{P_{aaa}\}$, $\mathcal{P}^{(4)}=\{P_{aaaa}, P_{aabb}, P_{abab}, P_{abba}\}$, 
%$\mathcal{P}^{(5)}=\{P_{aaaaa}, P_{aaabb}, P_{aabab}, P_{aabba},\ldots\}$, 
and so on.

\begin{table}[t]
\begin{center}
\begin{tabular}{|c||c|c|c|c|c|c|c|c|c|c|}
\hline
$n$ & 1 & 2 & 3 & 4 & 5 & 6 & 7 & 8 & 9 & 10  \\
\hline
$B_n$ & 1 & 2 & 5 & 15 & 52 & 203 & 877 & 4140 & 21147  & 115975 \\ 
%\hline
$F_n$ & 0 & 1 & 1 & 4 & 11 & 41 & 162 & 715 & 3425 & 17722 \\
$M_n(2)$ & 0 & 1 & 0 & 1 & 0 & 1 & 0 & 1 & 0 & 1    \\
$M_n(3)$ & 0 & 1 & 1 & 3 & 5 & 11 & 21 & 43 & 85 & 171    \\
$M_n(4)$ & 0 & 1 & 1 & 4 & 10 & 31 & 91 & 274 & 820 & 2461 \\
\hline
\end{tabular}
\caption{The Bell numbers $B_n$ give the number of partitions of $n$ points. The number $F_n$ of linearly independent $n$-point spin correlators (\ref{G}) coincides with the number of partitions on $n$ points into subsets containing at least two points. A number $M_n(q)$ of these correlators determines the $n$-point magnetic correlations in the $q$-state Potts ferromagnet.}
\label{tab_Bell}
\end{center}
\end{table}

\subsection{Spin correlators}
As we will see in a moment, it follows from the FK mapping that the Potts spin correlators  can be expressed as linear combinations of the cluster connectivities. Consistency of this statement requires that the number of linearly independent spin correlators coincides with the number of linearly independent cluster connectivities. The spin variables of the Potts model are defined as\footnote{Our present normalization of $\sigma_\alpha(x)$ differs by a factor $q$ from that used in \cite{DV2}.}
\EQ
\sigma_\alpha(x)=q\delta_{s(x),\alpha}-1\,,\hspace{1cm}\alpha=1,\ldots,q\,,
\label{spin}
\EN
where $s(x)$ is the color variable appearing in (\ref{H}), and satisfy
\EQ
\sum_{\alpha=1}^q\sigma_\alpha(x)=0\,.
\label{constraint}
\EN
The expectation value $\langle\sigma_\alpha\rangle$ is the order parameter of the Potts transition, since it differs from zero only in the spontaneously broken phase. More generally we denote by
\EQ
G_{\alpha_1\ldots\alpha_n}(x_1,\ldots,x_n)=\langle\sigma_{\alpha_1}(x_1)\ldots\sigma_{\alpha_n}(x_n)\rangle_{J\leq J_c}
\label{G}
\EN
the $n$-point spin correlators in the symmetric phase. We now show that, for $q$ real parameter, the number of linearly independent functions (\ref{G}) coincides with $F_n$. 

Let the string $(\alpha_1\ldots\alpha_n)$ identify the correlator (\ref{G}), and suppose that $\alpha_k$ is isolated  within this string, i.e. 
%$\alpha_k\not=\alpha_i$ $\forall i\not=k$
it is not fixed to coincide with any other index within the string. We can then use (\ref{constraint}) to sum over $\alpha_k$ and obtain, exploiting permutational symmetry, a linear relation among correlators involving a string similar to the original one together with strings without isolated indices. The simplest example,
\EQ
0=\sum_{\beta}G_{\alpha\beta}=G_{\alpha\alpha}+(q-1)G_{\alpha\gamma}\,,\hspace{1cm}\gamma\neq\alpha\,,
\label{example}
\EN
is sufficient to understand that (\ref{constraint}) produces meaningful equations also if $q$ is non-integer, the only consequence being that some multiplicity factors in front of the correlators become non-integer; also, the requirement $\gamma\neq\alpha$ should imply $q\geq 2$, but in the sense of analytic continuation to real values of $q$ (\ref{example}) is equally valid for $q<2$. Similarly, starting with a string with $m$ isolated indices $\alpha_{k_1},\ldots,\alpha_{k_m}$, summing over $\alpha_{k_1}$ and using permutational symmetry we generate a linear relation involving the original string with the $m$ isolated indices $\alpha_{k_1},\ldots,\alpha_{k_m}$ together with strings in which $\alpha_{k_1}$ is no more isolated, i.e. with at most\footnote{We can have strings with $m-1$ or $m-2$ isolated indices.} $m-1$ isolated indices. Iterating this procedure, any correlation function (\ref{G}) with isolated indices can be written as linear combination of correlation functions without isolated indices. Recalling the meaning of $F_n$ in terms of set partitions, we then see that there are at most $F_n$ $S_q$-inequivalent, linearly independent correlation functions (\ref{G}), i.e. those without isolated indices. On the other hand, the constraint (\ref{constraint}) does not generate any new linear relation if we start with a string which does not contain any isolated index. The number of linearly independent functions (\ref{G}) is then exactly $F_n$.  

Let us now detail the linear relation between cluster connectivities and spin correlators. The latter admit 
the FK clusters expansion,
\EQ
G_{\alpha_1\ldots\alpha_n}(x_1,\ldots,x_n)=\frac{1}{Z}\sideset{}{'}\sum_{s(x_1),\ldots,s(x_n)}\sum_{\mathcal{G}\subseteq
\mathcal{L}} p^{n_b}(1-p)^{\bar{n}_b}\prod_{i=1}^n\bigl(q\delta_{s(x_i),\alpha_i}-1\bigr)\,,
\label{FKcorr}
\EN
where the prime on the first sum means that sites belonging to the same FK cluster are forced to have the same color $s$. Notice that, if one of the points $x_i$ is isolated from the
others in a cluster of a given graph $\mathcal{G}$, then the sum over its colors gives zero due to (\ref{constraint}); hence, consistently with our previous analysis, (\ref{FKcorr}) receives a contribution only from partitions of the sites $x_i$'s into clusters containing at least two of these sites. If $P_{a_1\dots a_n}(x_1,\dots,x_n)$  is the probability of such a partition, then the number of distinct clusters will be $m<n$, and to any pair $(x_i,\alpha_i)$ we can  associate one of
the distinct letters $c_1,\dots,c_m$ chosen among the $a_i$'s. The coefficient of $P_{a_1\dots
  a_n}$ in the expansion (\ref{FKcorr}) is then\footnote{The prefactor $1/q^m$
ensures the correct probability measure for the graph $\mathcal{G}$.}
\EQ
\frac{1}{q^{m}}\sum_{s_1=1}^q\prod_{x_i\subset c_1}\bigl(q\delta_{s_1,\alpha_i}-1\bigr)\dots\sum_{s_{m}=1}^q\prod_{x_i\subset c_m}\bigl(q\delta_{s_{m},\alpha_i}-1\bigr);
\label{prob_coefficients}
\EN
the notation $x_i\subset c$ means that to the point $x_i$ is associated the
letter $c$ ($x$ belongs to the cluster $c$). For $n=2,3$ the dimensionality of correlation spaces is $F_2=F_3=1$ and (\ref{prob_coefficients}) gives
\bea
G_{\alpha\alpha} &=& q_1\,P_{aa}\,,
\label{G2}\\
G_{\alpha\alpha\alpha} &=& q_1q_2\,P_{aaa}\,,
\label{G3}
\eea
where we introduced the notation 
\EQ
q_k\equiv q-k\,.
\EN
The first relation with a matrix form appears at the four-point level ($F_4=4$), for which (\ref{prob_coefficients}) leads to
\begin{align}
&G_{\alpha\alpha\alpha\alpha}%\langle\sigma_{\alpha}\sigma_{\alpha}\sigma_{\alpha}\sigma_{\alpha}\rangle
=q_1(q^2-3q+3)P_{aaaa}+q_1^2(P_{aabb}+P_{abba}+P_{abab}),
\label{prob_aaaa}\\
&G_{\alpha\alpha\beta\beta}%\langle\sigma_{\alpha}\sigma_{\alpha}\sigma_{\beta}\sigma_{\beta}\rangle%
                           =(2q-3)P_{aaaa}+q_1^2P_{aabb}+P_{abba}+P_{abab},\label{prob_aabb}\\
&G_{\alpha\beta\beta\alpha}%\langle\sigma_{\alpha}\sigma_{\beta}\sigma_{\beta}\sigma_{\alpha}\rangle%
                           =(2q-3)P_{aaaa}+P_{aabb}+q_1^2P_{abba}+P_{abab},\label{prob_abba}\\
&G_{\alpha\beta\alpha\beta}%\langle\sigma_{\alpha}\sigma_{\beta}\sigma_{\alpha}\sigma_{\beta}\rangle%
                           =(2q-3)P_{aaaa}+P_{aabb}+P_{abba}+q_1^2P_{abab}.
\label{prob_abab}
\end{align}

\begin{figure}[t]
\begin{center}
\includegraphics[width=10.5cm]{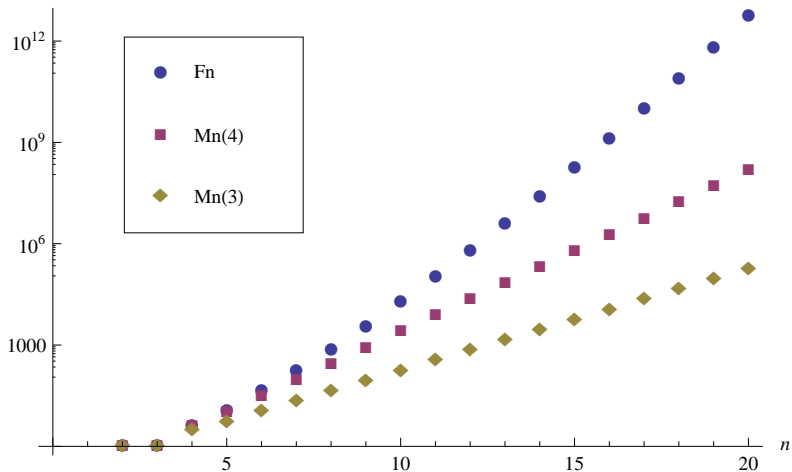}
\caption{Plots of the first 20 values of the sequences $F_n$ (number of independent $n$-point spin correlators for $q$ real) and $M_n(q)$ (number of independent $n$-point spin correlators in the magnetic sector) for $q=3,4$.}
\label{fig_asympt}
\end{center}
\end{figure}

The last set of equations, as well as those one obtains for $n>4$, can be inverted to express the connectivities in terms of the spin correlators, making clear that all the $F_n$ inequivalent and independent spin correlators are necessary to determine the connectivities of the random cluster model. On the other hand, a number $M_n(q)\leq F_n$ of these spin correlators determine the magnetic correlations in the Potts model at $q$ integer. This is due to the fact that the spin correlators are themselves the magnetic observables, and that for $q$ integer some of them vanish (e.g. $G_{\alpha\alpha\ldots\alpha}(x_1,\ldots,x_n)$ at $q=2$, $n$ odd, see (\ref{G3})), those involving more than $q$ colors are meaningless in the magnetic context, and additional linear relations may hold at specific values\footnote{For example the relation $3G_{\alpha\alpha\alpha\alpha}=2(G_{\alpha\beta\beta\alpha}+G_{\alpha\beta\alpha\beta}+G_{\alpha\alpha\beta\beta})$ holds specifically at $q=3$ and the system of equations (19)--(22) is no longer invertible. More generally, we expect that the $F_n\times F_n$ matrix $T_n(q)$ giving the spin correlators in terms of the \lq\lq basic\rq\rq\ connectivities (given explicitly by (\ref{G2}), (\ref{G3}) and (\ref{prob_aaaa})--(\ref{prob_abab}) for $n=2,3,4$) has determinant 
\EQ
\det(T_n)=q^{a_n}(q-1)\prod_{k=2}^{n-1}(q-k)^{d_n(k)}\,,
\label{det}
\EN
with $d_n(k)=\sum_{j=1}^k\tilde{S}(n,j)-M_n(k)$, $\tilde{S}(n,j)$ being the generalized Stirling numbers discussed in Appendix A, and $a_n$ determined by the requirement that the total degree of the polynomial (\ref{det}) is $D_n=\sum_{k=1}^n(n-k)\tilde{S}(n,k)$, as follows examining (\ref{prob_coefficients}). It follows from (\ref{F-S}) and (\ref{M-F}) that $d_n(k)=0$ for $k\geq n$. We checked (\ref{det}) up to $n=5$.} 
of $q$. The numerical sequences $M_n(q)$ are determined in Appendix~B for $q=2,3,4$ (the case $q=2$ is of course trivial); the first few values are given in Table~\ref{tab_Bell} and plots are shown in Fig.~\ref{fig_asympt}.
It is interesting, in particular,  to compare the large $n$ behavior of the dimensionalities of the magnetic and percolative correlation spaces. Defining
\EQ 
M_n(q)\stackrel{n\gg 1}{\sim}\text{e}^{n s_q(n)},
\EN
(\ref{rec_Mn_3}) and (\ref{rec_Mn_4}) give $s_q(n)=\log(q-1)$, a result which can be
consistently interpreted as a kind of entropy for the Potts spin $\sigma_{\alpha}$ and is expected to hold for any  integer $q>1$. $F_n$ exhibits instead the super-exponential growth \cite{Bruijn}
\EQ
F_n\stackrel{n\gg 1}{\sim}\text{e}^{n s(n)}\,,
\EN
\EQ
s(n)=\log n-\log\log n-1+\frac{\log\log n}{\log n}+O\left(\frac{1}{\log n}\right)\,.
\EN

\subsection{Scaling limit and correlators of kink fields}
For $q\leq q_c$, i.e. when the phase transition is continuous, the Potts field theory describes the scaling limit $J\to J_c$ of the Potts model, with the spin variables $\sigma_\alpha(x)$ playing the role of fundamental fields ($x$ is now a point in Euclidean space). In particular, the $q$ degenerate ferromagnetic ground states which the Potts model possesses above $J_c$ correspond in the scaling limit to degenerate vacua of the Potts field theory. In the two-dimensional case the kinks interpolating between a vacuum with color $\alpha$ and one with color $\beta$ are topologically stable and provide the elementary excitations of the spontaneously broken phase\footnote{See \cite{DJS} for a lattice study of Potts kinks.}; they are created by the kink fields $\mu_{\alpha\beta}(x)$, which are non-local with respect to the spin fields $\sigma_\alpha$. The products of kink fields are subject to the adjacency condition $\mu_{\alpha\beta}(x)\mu_{\beta\gamma}(y)$. Duality relates, in a way to be investigated in the next sections, the kink field correlators in the broken phase
\EQ
\tilde{G}_{\beta_1\ldots\beta_n}(x_1,\ldots,x_n)=\langle\mu_{\beta_1\beta_2}(x_1)\mu_{\beta_2\beta_3}(x_2)\ldots\mu_{\beta_n\beta_{1}}(x_n)\rangle_{J\geq J_c}
\label{Gtilde}
\EN
to the spin correlators in the symmetric phase (\ref{G}). Consistency of the duality relation requires that the number of $S_q$-inequivalent correlators (\ref{Gtilde}) coincides again with $F_n$. We now show that this is indeed the case.

Consider to start with the string of kink fields $\mu_{\beta_1\beta_2}\mu_{\beta_2\beta_3}\dots\mu_{\beta_n\beta_{n+1}}$ and associate to it $n+1$ points $P_i$, $i=1,\ldots,n+1$, on a line. Each point $P_i$ has a color $\beta_i$ which must differ from those of the adjacent points. Let us show first of all that the number $C_{n+1}$ of $S_q$-inequivalent colorations of the points $P_i$ coincides with the Bell number $B_n$. If the adjacency condition is relaxed, the number of inequivalent colorations of the $n+1$ points is $B_{n+1}$. The string will consist of $k+1$ substrings, each with a definite color different from those of the adjacent substrings, that we can think to separate by placing $k$ domain walls between them; this can be done in $\binom{n}{k}$ ways. The $k+1$ substrings can be colored in $C_{k+1}$ $S_q$-inequivalent ways and we have
\EQ
B_{n+1}=\sum_{k=0}^{n}\binom{n}{k}C_{k+1}\quad\forall n\geq 0. 
\label{Bell_recursion}
\EN
Since $C_1=1$, the result $C_{n+1}=B_n$ then follows from (\ref{bell_def}) by induction. 

The case we just discussed includes ${L}_{n+1}$ inequivalent colorations in which $\beta_1=\beta_{n+1}$ (the case (\ref{Gtilde}) we are actually interested in) and $O_{n+1}$ inequivalent colorations in which $\beta_1\neq\beta_{n+1}$, i.e. ${L}_{n+1}+O_{n+1}=C_{n+1}$. On the other hand, if we start with $n$ points having $\beta_1\neq\beta_n$ and we add $P_{n+1}$ with $\beta_{n+1}=\beta_1$, the number of inequivalent colorations does not change, i.e. ${L}_{n+1}=O_n$. We then see that
\EQ
{L}_{n+1}+{L}_{n+2}=C_{n+1}=B_{n}\,,\hspace{1cm}\forall n\geq 0\,.
\label{LCB}
\EN
This relation, together with the initial condition ${L}_1=1$, can be used to generate the ${L}_n$'s from the $B_n$'s. Of course ${L}_{n+1}$ is the number of inequivalent correlators (\ref{Gtilde}) we were looking for.  Comparison of (\ref{LCB}) with (\ref{Fdef}) then leads to the final identification $L_{n+1}=F_n$.

\section{Operator product expansions}
Generically the OPE of scalar fields $A_i(x)$ with scaling dimension $X_i$ takes the form
\EQ
\lim_{x_1\rightarrow x_2} A_i(x_1) A_j(x_2)=\sum_{m}C_{ij}^m\frac{A_m(x_2)}{x_{12}^{X_i+X_j-X_m}}\,,
\label{OPE_coeff}
\EN
where we include for simplicity only scalar fields in the r.h.s. and use the notation $x_{ij}\equiv|x_i-x_j|$; in the following we will replace (\ref{OPE_coeff}) by the symbolic notation
\EQ
A_i\cdot A_j=\sum_{m}C_{ij}^m\,A_m\,.
\EN
The nature of the fields $\mu_{\alpha\beta}(x)$ naturally leads to the two-channel OPE \cite{DV2}
\EQ
\mu_{\alpha\beta}\cdot\mu_{\beta\gamma}=\delta_{\alpha\gamma}\tilde{I}+(1-\delta_{\alpha\gamma})(C_{\mu}\mu_{\alpha\gamma}+\ldots)\,,
\label{kink_OPE}
\EN  
where the neutral channel $\alpha=\gamma$ contains the expansion $\tilde{I}=I+C_{\varepsilon}\varepsilon+\ldots$ over $S_q$-invariant fields (identity $I$, energy $\varepsilon$, and so on), and the charged channel $\alpha\neq\gamma$ the expansion over $\mu_{\alpha\gamma}$ and less relevant kink fields; $C_\varepsilon$ and $C_\mu$ are simplified notations for the OPE coefficients, for which exact expressions for continuous $q$ have been given in \cite{DV2}. 

The fields $\mu_{\alpha\beta}(x)$ are expected to be related to the disorder fields $\mu_\alpha(x)$ by the linear transformation
\EQ
\mu_{\alpha}(x)=\sum_{\sigma}C_{\alpha}^{\rho\sigma}\mu_{\rho\sigma}(x)\,,
\label{mapping}
\EN 
where $C_{\alpha}^{\rho\sigma}\in\mathbb C$ are coefficients\footnote{No confusion should arise with the OPE coefficients $C_{ij}^m$ of (\ref{OPE_coeff}).} to be investigated below, and $\rho$-independence is a consequence of permutational symmetry. The field $\mu_{\alpha}$ carries the same representation of permutational symmetry as $\sigma_\alpha$ (in particular, $\sum_{\alpha=1}^q\mu_\alpha=0$) but, as $\mu_{\alpha\beta}$, it is non-local with respect to $\sigma_\alpha$. In other words, $\sigma_\alpha$ and $\mu_\alpha$ are identical (dual) fields living in mutually non-local sectors of the theory; in particular they share the same scaling dimension and the same OPE. Mutual non-locality reflects in the fact that, while $\langle\sigma_\alpha\rangle\neq 0$ for $J>J_c$, $\langle\mu_\alpha\rangle\neq 0$ for $J<J_c$. More precisely, in view of the coinciding scaling dimension, it is sufficient to adopt the same normalization of the fields to ensure that $\langle\sigma_\alpha\rangle_J=\langle\mu_\alpha\rangle_{J^*}$, where $J^*$ is the dual\footnote{In the scaling limit we consider, $J$ and $J^*$ are the points where the elementary excitations of the symmetric phase and those of the spontaneously broken phase have the same mass $m$; $m=0$ at the self-dual point $J_c$.} of $J$. The duality extends to multi-point functions, in such a way that the spin correlators (\ref{G}) can also be written as
\EQ
G_{\alpha_1\ldots\alpha_n}(x_1,\ldots,x_n)=\langle\mu_{\alpha_1}(x_1)\ldots\mu_{\alpha_n}(x_n)\rangle_{J^*\geq J_c}\,.
\label{Gdual}
\EN
The rest of this section is devoted to investigate the relation (\ref{mapping}) and to determine the structure of the OPE $\mu_\alpha\cdot\mu_\beta$ (or, equivalently, $\sigma_\alpha\cdot\sigma_\beta$).

\begin{figure}[t]
\begin{center}
\includegraphics[width=11cm]{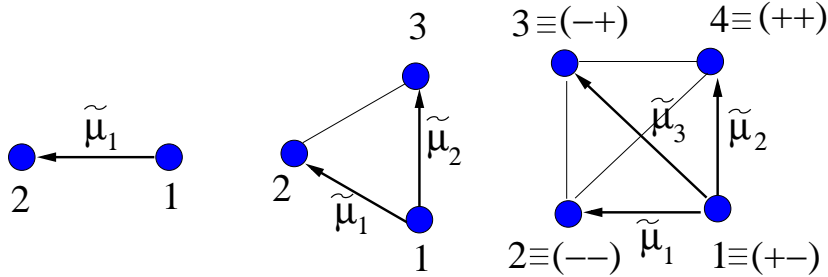}
\caption{The vacua of the Potts field theory for $q=2,3,4$ are labeled by
  $\alpha=1,\ldots,q$ and denoted by dots (for $q=4$ we use the Ashkin-Teller notation $\boldsymbol{\alpha}\equiv(\alpha_1, \alpha_2)$). Fixing a specific vacuum (1 in this case) amounts to choose a representative kink field within each class $\tilde{\mu}_i$ (see the text).}
\label{fig_vuoti}
\end{center}
\end{figure}

The $q$ degenerate vacua of the Potts model above $J_c$ can be associated to the vertices of an hypertetrahedron in $q-1$ dimensions whose $q(q-1)$ oriented sides are associated to the kink fields $\mu_{\alpha\beta}$. Permutational symmetry of the vacua allows to group these fields into classes $\tilde{\mu}_i$, $i=1,\ldots,q-1$, each containing $q$ kink fields starting from different vacua, in such a way that choosing a vacuum amounts to select $q-1$ kink fields, one from each class, starting from that vacuum and arriving at the other vacua (Fig.~\ref{fig_vuoti}). Ignoring structure constants, the OPE (31) has the form of a multiplication between elements of a finite group. Independence from the choice of the  starting vacuum of the kinks ensures that the elements of this finite group are  the classes $\tilde{\mu}_i$, $i=1,\dots, q-1$, together with the topologically neutral class $\tilde{I}$. We denote then by $K_q$ their fusion table as prescribed by (31), as well as the finite group of order $q$ it defines. The symmetry also ensures that all the rows of the matrix $C_{\alpha}$ can be obtained from the first by regular permutations\footnote{Permutations which do not leave any element invariant. It is not difficult to realize that such permutations are elements of $K_q$.}. The relation (\ref{mapping}) (which we could equivalently write as  $\mu_{\alpha}=\sum_{\rho}C_{\alpha}^{\rho\sigma}\mu_{\rho\sigma}$) is effectively a sum over the $q-1$ classes $\tilde{\mu}_i$, as we now illustrate separately discussing the cases $q=2,3,4$.

\noindent
{\bf q=2.} Permutational symmetry $S_2=\mathbb Z_2$ implies the equivalence of the two
kink fields $\mu_{12}$ and $\mu_{21}$, which we collect in the class $\tilde{\mu}_1$; from (\ref{kink_OPE}) we derive $K_2=\mathbb Z_2$ (see Fig.~\ref{fig_tabelle}). We can choose
\EQ
\mu_{\alpha}=\omega_2^{\alpha}\tilde{\mu}_1,\quad
C_{\alpha}=\left(\begin{array}{ccc}0 & \omega_2^{\alpha}
  \\  \omega_2^{\alpha}& 0\end{array}\right),\quad\alpha=1,2\,,
\label{mapping_q2}
\EN
with $\omega_q=\text{e}^{2\pi i/q}$, in order to fulfill
$\sum_{\alpha}\mu_{\alpha}=0$ and consistently derive
\begin{align}
\mu_{\alpha}\cdot\mu_{\alpha}&=\tilde{I},\label{aa2}\\
\mu_{\alpha}\cdot\mu_{\beta}&=-\tilde{I}\,,\hspace{1cm}\alpha\neq\beta\,.
\label{muOPE_q2}
\end{align}

\noindent
{\bf q=3.} The OPE of the kink fields $\mu_{\alpha\beta}$ is equivalent to the fusion table of the
classes $\tilde{\mu}_1, \tilde{\mu}_2$ and $\tilde{I}$. Being $\mathbb{Z}_3$ the only discrete group of order three, full consistency requires $K_3=\mathbb Z_3$, together with the
identifications $\tilde{\mu}_1=\{\mu_{12}, \mu_{23}, \mu_{31}\},\text{ }\tilde{\mu}_2=\{\mu_{13},
\mu_{21}, \mu_{32}\}$. Notice that $\text{Aut}(\mathbb Z_3)=\mathbb Z_2$, and the non-trivial 
automorphism\footnote{Given a group $G$, $\phi: G\rightarrow G$ is an
  automorphism if $\phi(ab)=\phi(a)\phi(b), \forall a,b\in G$. The set of all
  the automorphisms with natural composition as a product form a group called
  $\text{Aut}(G)$. } corresponds to the charge conjugation operator
$\mathcal{C}$, with $\mathcal{C}\tilde{\mu}_1=\tilde{\mu}_2$. Using $\mu_{\alpha}(x)=3\delta_{\tilde{s}(x),\alpha}-1$, the charge conjugated operators realizing the $\mathbb Z_3$ OPE are identified with\footnote{The basis $\tilde{\mu}_1$, $\tilde{\mu}_2$ is that used in \cite{ZF}.} $\tilde{\mu}_1=\text{e}^{2\pi i \tilde{s}(x)/3}$ and $\tilde{\mu}_2=\text{e}^{-2\pi i \tilde{s}(x)/3}$, where $\tilde{s}(x)$ is the dual color variable; using also $\delta_{\tilde{s},\alpha}=\frac{1}{3}\sum_{\beta=1}^3 \text{e}^{\frac{2\pi i}{3} (\tilde{s}-\alpha)\beta}$ one obtains
\EQ
\mu_{\alpha}=\omega_3^{-\alpha}\tilde{\mu}_1+\omega_3^{\alpha}\tilde{\mu}_2,\quad
C_{\alpha}=\left(\begin{array}{ccc} 0 & \omega_3^{-\alpha} & \omega_3^{\alpha} \\
\omega_3^{\alpha}& 0 & \omega_3^{-\alpha} \\
\omega_3^{-\alpha}& \omega_3^{\alpha} & 0 
\end{array}\right),\quad\alpha=1,2,3\,,
\label{mapping_q3}
\EN
and then
\begin{align}
\mu_{\alpha}\cdot\mu_{\alpha}&=2\tilde{I}+C_{\mu}\,\mu_{\alpha}+\ldots,\label{aa3}\\
\mu_{\alpha}\cdot\mu_{\beta}&=-\tilde{I}-C_{\mu}(\mu_{\alpha}+\mu_{\beta})+\ldots\,,\hspace{1cm}\alpha\neq\beta\,;
\label{muOPE_q3}
\end{align} 
the relation $\omega_3^{\alpha}+\omega_3^{\beta}+\omega_3^{-(\alpha+\beta)}=0$, $\alpha\neq\beta$, is used.

\begin{figure}[t]
\begin{center}
\begin{tabular}{c|cc}
$\cdot$& $\tilde{I}$ & $\tilde{\mu}_1$ \\
\hline
$\tilde{I}$& $\tilde{I}$ & $\tilde{\mu_1}$ \\
$\tilde{\mu_1}$& $\tilde{\mu}_1$& $\tilde{I}$ 
\end{tabular}\qquad
\begin{tabular}{c|ccc}
$\cdot$& $\tilde{I}$ & $\tilde{\mu}_1$& $\tilde{\mu}_2$ \\
\hline
$\tilde{I}$& $\tilde{I}$ & $\tilde{\mu_1}$& $\tilde{\mu}_2$ \\
$\tilde{\mu_1}$& $\tilde{\mu}_1$& $\tilde{\mu}_2$& $\tilde{I}$\\
$\tilde{\mu_2}$& $\tilde{\mu}_2$& $\tilde{I}$& $\tilde{\mu}_1$ 
\end{tabular}\qquad
\begin{tabular}{c|cccc}
$\cdot$& $\tilde{I}$ & $\tilde{\mu}_1$& $\tilde{\mu}_2$& $\tilde{\mu}_3$ \\
\hline
$\tilde{I}$& $\tilde{I}$ & $\tilde{\mu_1}$& $\tilde{\mu}_2$ &$\tilde{\mu}_3$\\
$\tilde{\mu_1}$& $\tilde{\mu}_1$& $\tilde{I}$& $\tilde{\mu_3}$&$\tilde{\mu}_2$ \\
$\tilde{\mu_2}$& $\tilde{\mu}_2$& $\tilde{\mu}_3$& $\tilde{I}$&$\tilde{\mu}_1$ \\
$\tilde{\mu_3}$& $\tilde{\mu}_3$&$\tilde{\mu_2}$& $\tilde{\mu}_1$&$\tilde{I}$ 
\end{tabular}
\caption{Fusion tables $K_q$ at $q=2,3,4$. They correspond to the groups $\mathbb{Z}_2$, $\mathbb{Z}_3$ and $D_2$, respectively.}
\label{fig_tabelle}
\end{center}
\end{figure}

\noindent
{\bf q=4.} The four-state Potts model can be seen as the case $J=J_4$ of the Ashkin-Teller model
defined by the Hamiltonian
\EQ
{\cal H}_{AT}=-\sum_{\langle x,y\rangle}\{J[\tau_1(x)\tau_1(y)+\tau_2(x)\tau_2(y)]+J_4\,\tau_1(x)\tau_1(y)\tau_2(x)\tau_2(y)\},
\label{AT}
\EN
where $\tau_i=\pm 1$, $i=1,2$, are Ising variables. Defining $\boldsymbol{s}=(\tau_1, \tau_2)$, $\boldsymbol{\alpha}=(\alpha_1, \alpha_2)$, with $\alpha_i=\pm 1$, and
$\delta_{\boldsymbol{s},\boldsymbol{\alpha}}=\delta_{\tau_1,\alpha_1}\delta_{\tau_2,\alpha_2}$, the Potts spin (\ref{spin}) can be written as
\EQ
\sigma_{\boldsymbol{\alpha}}=4\delta_{\boldsymbol{s},\boldsymbol{\alpha}}-1=\alpha_1\tau_1+\alpha_2\tau_2+\alpha_1\alpha_2\tau_1\tau_2.
\label{4Potts_2Ising}
\EN
The kink fields $\mu_{\boldsymbol{\alpha}\boldsymbol{\beta}}$ interpolate between the four degenerate vacua of the two coupled Ising models (see e.g. \cite{DG}); the classes $\tilde{\mu}_1=\{\mu_{12},\mu_{21},\mu_{34},\mu_{43}\}$ and $\tilde{\mu}_2=\{\mu_{14},\mu_{41},\mu_{23},\mu_{32}\}$ are constructed in analogy to the case $q=2$, the fields in $\tilde{\mu}_3=\{\mu_{13}, \mu_{31}, \mu_{24}, \mu_{42}\}$ are instead obtained taking the OPE according to (31) (see also Fig. 2). We derive $K_4=D_2=\mathbb{Z}_2\times\mathbb{Z}_2$, see
Fig.~\ref{fig_tabelle}; notice that $\text{Aut}(D_2)=S_3$. We can also take
\EQ
\mu_{\boldsymbol{\alpha}}=\alpha_1\tilde{\mu}_1+\alpha_2\tilde{\mu}_2+\alpha_1\alpha_2\tilde{\mu}_3\,,\quad\quad C_{\boldsymbol{\alpha}}=\left(\begin{array}{cccc} 0 & \alpha_1 & \alpha_1\alpha_2&\alpha_2 \\
\alpha_1& 0 & \alpha_2& \alpha_1\alpha_2 \\
\alpha_1\alpha_2& \alpha_2 & 0 & \alpha_1 \\
\alpha_2& \alpha_1\alpha_2 &\alpha_1 & 0 
\end{array}\right)\,,
\label{mapping_q4} 
\EN
from which we obtain
\begin{align}
\mu_{\boldsymbol{\alpha}}\cdot\mu_{\boldsymbol{\alpha}}&=3\tilde{I}+2C_{\mu}\,\mu_{\boldsymbol{\alpha}}+\ldots\,,\label{aa4}\\
\mu_{\boldsymbol{\alpha}}\cdot\mu_{\boldsymbol{\beta}}&=-\tilde{I}-C_{\mu}(\mu_{\boldsymbol{\alpha}}+\mu_{\boldsymbol{\beta}})+\dots\,,\hspace{1cm}\alpha\neq\beta\,.
\label{muOPE_q4}
\end{align}

It is interesting to remark some formal properties emerging from this analysis. We see that, by construction, $K_q$ at $q=2,3,4$ is a finite abelian group of order $q$, i.e. by Cayley theorem a regular abelian subgroup\footnote{The classes $\tilde{\mu}_i$ are
  associated to the regular permutations $\pi_i$, $i=1,\dots,q-1$, of $S_q$ as  $\tilde{\mu}_i=\{\mu_{1\pi_i(1)},...,\mu_{q\pi_i(q)}\}$. Without loss of
  generality one can assume $\pi_i(1)=i+1$.} of $S_q$. $K_q$ must also be invariant under permutations of the $q-1$ classes $\tilde{\mu}_i$, an operation which
corresponds to fix one vacuum and permute the remaining $q-1$. Formally this amounts to write $\text{Aut}(K_q)=S_{q-1}$, and we expect the full symmetry group of the theory to be realized as\footnote{The presence of the semidirect product $\rtimes$ is due to the fact that $S_{q-1}$ is not a normal subgroup of $S_q$.} 
\EQ
S_q=K_q\rtimes S_{q-1}\,.
\label{semidirect}
\EN 
This in turn implies the possibility of writing $S_q$ as a semidirect product of abelian subgroups of the form
\EQ
S_q=K_q\rtimes K_{q-1}\rtimes\dots\rtimes K_2,\quad K_2=\mathbb Z_2,
\label{solvability}
\EN 
a property which is equivalent to the solvability of the permutational
group. More precisely\footnote{We thank C. Casolo for this observation. Interesting remarks about solvable groups and lattice duality can be found in \cite{DIZ}.}, the solvability of $S_q$ would imply the existence of the factorization  (\ref{solvability}), as indeed remarkably happens at $q=2,3,4$, with $K_2=\mathbb{Z}_2$, $K_3=\mathbb{Z}_3$, $K_4=D_2$ and
$\text{Aut}(\mathbb{Z}_3)=\mathbb{Z}_2$, $\text{Aut}(D_2)=S_3$. It is well known \cite{H}, however, that  for $q>4$ $S_q$ possesses no abelian normal subgroup, making impossible in particular to realize the condition
(\ref{semidirect}). Then $S_q$ is  not solvable for any
$q>4$, a circumstance which is interesting to compare with the fact that $q_c=4$ is also the upper bound for the existence of the Potts field theory in the two-dimensional case, i.e. the only case for which kink fields exist and the above construction is possible.

A remarkable feature appearing from (\ref{muOPE_q2}), (\ref{muOPE_q3}) and (\ref{muOPE_q4}) is that $\mu_\alpha\cdot\mu_{\beta\neq\alpha}$ is identical at $q=2,3,4$ (recall that $\mu_1+\mu_2=0$ at $q=2$). It is then absolutely natural to assume that this form actually holds unchanged for continuous values of $q$, and to write
\EQ
\sigma_{\alpha}\cdot\sigma_{\beta}=-I-C_{\mu}(\sigma_{\alpha}+\sigma_{\beta})+\ldots\,,\hspace{1cm}\alpha\neq\beta,
\label{OPE_sigma_ab}
\EN 
where the dots correspond to less relevant fields and we switched to the equivalent expression in terms of the spin fields for a reason to be made immediately clear. On the other hand, the complementary relation
\EQ
\sigma_{\alpha}\cdot\sigma_{\alpha}=q_1\,I+q_2C_{\mu}\,\sigma_{\alpha}+\cdots\
\label{OPE_sigma_aa}
\EN
follows observing that (\ref{constraint}) and (\ref{OPE_sigma_ab}) give
\begin{align}
0=\sigma_{\alpha}\cdot\sum_{\beta}\sigma_{\beta}&=
\sigma_{\alpha}\cdot\sigma_{\alpha}+\sum_{\beta\not=\alpha}\sigma_{\alpha}\cdot\sigma_{\beta}\nonumber\\
&=\sigma_{\alpha}\cdot\sigma_{\alpha}+\sum_{\beta\not=\alpha}\left[-I-C_{\mu}(\sigma_{\alpha}+\sigma_{\beta})+\ldots\right]\,;\nonumber
\end{align}
(\ref{OPE_sigma_aa}) is of course consistent with (\ref{aa2}), (\ref{aa3}) and (\ref{aa4}). While the disorder fields $\mu_\alpha(x)$ are specific of the two-dimensional case, the spin fields $\sigma_\alpha(x)$ are well defined in any dimension. It is then quite obvious to expect that (\ref{OPE_sigma_ab}) and (\ref{OPE_sigma_aa}) hold for real values of $q\leq q_c$ in any dimension.

The linear relation (\ref{constraint}) among the spin fields induces a relation less direct than usual between the OPE coefficients and the structure constants appearing in the three-point functions. In general, the structure constants $C_{ijk}$ are defined by the critical correlators
\EQ
\langle A_i(x_1)A_j(x_2)A_k(x_3)\rangle=\frac{C_{ijk}}{x_{12}^{X_i+X_j-X_k}x_{13}^{X_i+X_k-X_j}x_{23}^{X_j+X_k-X_i}}\,;
\label{struc_const}
\EN  
taking the limits $x_{12}\rightarrow 0$ and $x_{23}\rightarrow 0$ and using (\ref{OPE_coeff}), $C_{ijk}$ is expressed in term of the OPE coefficients as 
\EQ
C_{ijk}=\sum_{X_m=X_k}C_{ij}^mC_{mk}^I\,.
\EN 
We then find
\begin{align}
&C_{\sigma_{\alpha}\sigma_{\alpha}\sigma_{\alpha}}=q_2q_1C_{\mu},\\
&C_{\sigma_{\alpha}\sigma_{\alpha}\sigma_{\beta}}=C_{\sigma_{\alpha}\sigma_{\beta}\sigma_{\alpha}}=C_{\sigma_{\beta}\sigma_{\alpha}\sigma_{\alpha}}=-q_2C_{\mu},\\
&C_{\sigma_{\alpha}\sigma_{\beta}\sigma_{\gamma}}=2C_{\mu}\,,
\end{align}
with different indices denoting different colors.

\section{Duality relations}
Equations (\ref{mapping}) and (\ref{Gdual}) imply a linear relation between the spin correlators (\ref{G}) in the symmetric phase and the correlators (\ref{Gtilde}) of kink fields in the broken phase. This duality takes the form
\begin{align}
 G_{\alpha_1\dots\alpha_n}(x_1,\dots,x_n)&=
 \sum_{\beta_1,\dots,\beta_{n}}\tilde{D}_{\alpha_1\dots\alpha_n}^{\beta_1\dots\beta_{n}}\,\tilde{G}_{\beta_1\dots\beta_{n}}(x_1,\dots,x_n)\nonumber\\
&=\sideset{}{'}\sum_{\beta_1,\dots,\beta_{n}}\Bigl(\prod_{i=1}^{n_c\left(\boldsymbol{\beta}\right)-1}q_i\Bigr)\text{
 }D_{\alpha_1\dots\alpha_n}^{\beta_1\dots\beta_{n}}\,\tilde{G}_{\beta_1\dots\beta_{n}}(x_1,\dots,x_n)\,,
\label{duality_mapping}
\end{align} 
where the primed sum runs over all choices of $\boldsymbol{\beta}=\{\beta_1,\ldots,\beta_n\}$ which are inequivalent under permutations, $n_c(\boldsymbol{\beta})$ is the number of different colors in $\boldsymbol{\beta}$, and the factors $q_i$ have been extracted for later convenience. The task is that of determining the coefficients $D_{\alpha_1\dots\alpha_n}^{\beta_1\dots\beta_{n}}$, for continuous values of $q$; of course it is sufficient to consider a set of $F_n$ linearly independent spin correlators. We will discuss explicitly this problem up to the first case with $F_n>1$, i.e. $n=4$.

\noindent
{\bf n=1.} The trivial identity $\langle\sigma_{\alpha}\rangle=\langle\mu_{\alpha\beta}\rangle=0$ simply reflects the fact that we consider spin correlators in the symmetric phase\footnote{Since this is understood, here and in the following we omit the subscripts $J\leq J_c$ for spin correlators and $J\geq J_c$ for kink field correlators.}, and that $\mu_{\alpha\beta}$ is a kink field.

\noindent
{\bf n=2.} $F_2=1$ and we consider
\EQ
\langle\sigma_{\alpha}(x_1)\sigma_{\alpha}(x_2)\rangle=q_1D_{\alpha\alpha}^{\alpha\beta}\text{
}\langle\mu_{\alpha\beta}(x_1)\mu_{\beta\alpha}(x_2)\rangle\,.
\label{duality_2pts}
\EN
We can take the limit $x_{12}\rightarrow 0$ on both sides and equate the coefficients of the leading singularity $x_{12}^{-2X_\sigma}$, which are obtained using (\ref{OPE_sigma_aa}) and (\ref{kink_OPE}). This immediately yields $D_{\alpha\alpha}^{\alpha\beta}=1$. 

\noindent
{\bf n=3.} $F_3=1$ and we consider
\EQ
\langle\sigma_{\alpha}(x_1)\sigma_{\alpha}(x_2)\sigma_{\alpha}(x_3)\rangle=q_1q_2D_{\alpha\alpha\alpha}^{\alpha\beta\gamma}\text{
}\langle\mu_{\alpha\beta}(x_1)\mu_{\beta\gamma}(x_2)\mu_{\gamma\alpha}(x_3)\rangle\,.
\label{duality_3pts}
\EN   
We can again use the OPE's to take the limit $x_{12}\rightarrow 0$ on both sides and reduce\footnote{Notice that the OPE on the l.h.s. apparently produces singularities from the $S_q$-invariant operators $\mathcal{O}_k$ in $\tilde{I}$ which do not arise in the r.h.s., where $\alpha\neq\gamma$. Everything is consistent, however, since $\langle\mathcal{O}_k\sigma_{\alpha}\rangle=0$ by symmetry.} to (\ref{duality_2pts}); this leads to $D_{\alpha\alpha\alpha}^{\alpha\beta\gamma}=1$.

\begin{figure}[t]
\begin{center}
\includegraphics[width=9cm]{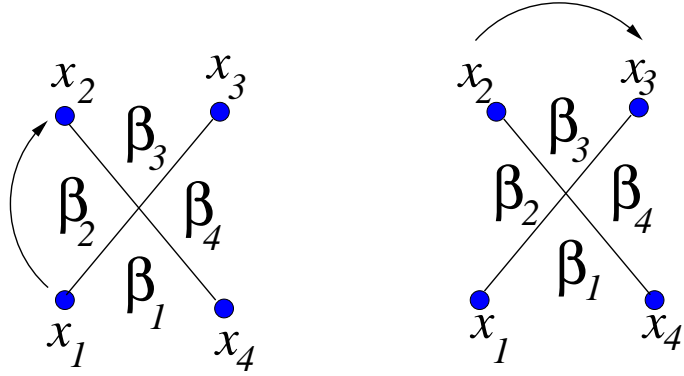}
\caption{Pictorial representation of the four-point correlation functions of kink fields $\tilde{G}_{\beta_1\beta_2\beta_3\beta_4}(x_1,\ldots,x_4)$. The two pinchings $x_1\rightarrow x_2$ and $x_2\rightarrow x_3$ used for the study of the duality relation (\ref{duality_4pts}) are indicated.}
\end{center}
\label{pinching}
\end{figure}

\noindent
{\bf n=4.} We will consider the $F_4=4$ linearly independent spin correlators (\ref{prob_aaaa}--\ref{prob_abab}), which expand as
\EQ
\begin{split}
\langle\sigma_{\alpha_1}(x_1)\sigma_{\alpha_2}(x_2)\sigma_{\alpha_3}(x_3)\sigma_{\alpha_4}(x_4)\rangle=q_1q_2q_3D_{\alpha_1\alpha_2\alpha_3\alpha_4}^{\alpha\beta\gamma\delta}\langle\mu_{\alpha\beta}(x_1)\mu_{\beta\gamma}(x_2)\mu_{\gamma\delta}(x_3)\mu_{\delta\alpha}(x_4)\rangle &\\
+q_1q_2
D_{\alpha_1\alpha_2\alpha_3\alpha_4}^{\alpha\beta\gamma\beta}\text{  }\langle\mu_{\alpha\beta}(x_1)\mu_{\beta\gamma}(x_2)\mu_{\gamma\beta}(x_3)\mu_{\beta\alpha}(x_4)\rangle &\\
+q_1q_2
D_{\alpha_1\alpha_2\alpha_3\alpha_4}^{\alpha\beta\alpha\gamma}\text{  }\langle\mu_{\alpha\beta}(x_1)\mu_{\beta\alpha}(x_2)\mu_{\alpha\gamma}(x_3)\mu_{\gamma\alpha}(x_4)\rangle &\\
+q_1
D_{\alpha_1\alpha_2\alpha_3\alpha_4}^{\alpha\beta\alpha\beta}\text{
}\langle\mu_{\alpha\beta}(x_1)\mu_{\beta\alpha}(x_2)\mu_{\alpha\beta}(x_3)\mu_{\beta\alpha}(x_4)\rangle &\,.\\
\label{duality_4pts}
\end{split}
\EN
We start again equating the coefficients of the short distance singularities on the two sides. However, an important difference with the cases $n<4$ now appears. Indeed, while the OPE in the l.h.s. can be taken for any pair of points $x_i$ and $x_j$, the kink nature of the fields on the r.h.s. allows us to use (\ref{kink_OPE}) only for adjacent fields. In other words, we can compare only the singularities arising for $x_{12}\rightarrow 0$ and $x_{23}\rightarrow 0$ (see Fig.~\ref{pinching}; $x_{34}\rightarrow 0$ and $x_{14}\rightarrow 0$ add nothing new). Equating the coefficients of the singularities in the ``neutral'' and ``charged'' channels, and using the duality relations already obtained for $n=2,3$, leads to the following sets of  equations 
\begin{align}
&\left\{\begin{aligned}
&D_{\alpha\alpha\alpha\alpha}^{\alpha\beta\alpha\gamma}=D_{\alpha\alpha\alpha\alpha}^{\alpha\beta\gamma\beta}\\
&q_2=q_3D_{\alpha\alpha\alpha\alpha}^{\alpha\beta\gamma\delta}+D_{\alpha\alpha\alpha\alpha}^{\alpha\beta\gamma\beta}\\
&q_1=q_2D_{\alpha\alpha\alpha\alpha}^{\alpha\beta\gamma\beta}+D_{\alpha\alpha\alpha\alpha}^{\alpha\beta\alpha\beta}
\end{aligned}\right.
&\left\{\begin{aligned}
&D_{\alpha\beta\alpha\beta}^{\alpha\beta\alpha\gamma}=D_{\alpha\beta\alpha\beta}^{\alpha\beta\gamma\beta}\\
&2=q_1q_3D_{\alpha\beta\alpha\beta}^{\alpha\beta\gamma\delta}+q_1D_{\alpha\beta\alpha\beta}^{\alpha\beta\gamma\beta}\\
&1=q_1q_2D_{\alpha\beta\alpha\beta}^{\alpha\beta\gamma\beta}+q_1D_{\alpha\beta\alpha\beta}^{\alpha\beta\alpha\beta}
\end{aligned}\right.
\label{4pts_OPE1}
\\ \nonumber\\
&\left\{\begin{aligned}
&1=q_1q_2D_{\alpha\alpha\beta\beta}^{\alpha\beta\gamma\beta}+q_1D_{\alpha\alpha\beta\beta}^{\alpha\beta\alpha\beta}\\
  &2=q_1q_3D_{\alpha\alpha\beta\beta}^{\alpha\beta\gamma\delta}+q_1D_{\alpha\alpha\beta\beta}^{\alpha\beta\alpha\gamma}\\
&q_1=q_2D_{\alpha\alpha\beta\beta}^{\alpha\beta\alpha\gamma}+D_{\alpha\alpha\beta\beta}^{\alpha\beta\alpha\beta}
\end{aligned}\right.&\left\{\begin{aligned}
&1=q_1q_2D_{\alpha\beta\beta\alpha}^{\alpha\beta\alpha\gamma}+q_1D_{\alpha\beta\beta\alpha}^{\alpha\beta\alpha\beta}\\
  &2=q_1q_3D_{\alpha\beta\beta\alpha}^{\alpha\beta\gamma\delta}+q_1D_{\alpha\beta\beta\alpha}^{\alpha\beta\gamma\beta}\\
&q_1=q_2D_{\alpha\beta\beta\alpha}^{\alpha\beta\gamma\beta}+D_{\alpha\beta\beta\alpha}^{\alpha\beta\alpha\beta}
\end{aligned}\right.
\label{4pts_OPE2}
\end{align}
where different indices denote different colors.
Notice that the method produces four equations for each correlation function (two per pinching and per channel), but only three turn out to be independent. Since each duality relation (\ref{duality_4pts}) involves four coefficients, the above equations are not sufficient to fix everything. We now show how the duality for the correlators $G_{\alpha\alpha\alpha\alpha}$, $G_{\alpha\alpha\beta\beta}$, $G_{\alpha\beta\beta\alpha}$ can be completely determined exploiting also the relations (\ref{mapping}), (\ref{Gdual}).

The matrices $C_{\alpha}$ defined in (\ref{mapping}) are hermitian, due to the presence of the antilinear charge conjugation operator
$\mathcal{C}$, with $\mathcal{C}^2=I$, 
$\mathcal{C}\mu_{\rho\sigma}\mathcal{C}=\mu_{\sigma\rho}$, 
$\mathcal{C}\mu_{\alpha}\mathcal{C}=\mu_{\alpha}$ and $\mathcal{C}C_{\alpha}^{\rho\sigma}\mathcal{C}=\left(C_{\alpha}^{\rho\sigma}\right)^{*}$. They also satisfy $C_{\alpha}^{\rho\rho}=0$, $\rho=1,\dots,q$, and $\sum_{\alpha}C_{\alpha}=0$. Other properties follow requiring the consistency of the OPE's for the kink fields and for the dual spin $\mu_{\alpha}$. For example,
\begin{align}
\mu_{\alpha}\cdot\mu_{\alpha}&=\sum_{\sigma,\omega}C_{\alpha}^{\rho\sigma}C_{\alpha}^{\sigma\omega}\mu_{\rho\sigma}\cdot\mu_{\sigma\omega}\\
&=\biggl(\sum_{\sigma\not=\rho}C_{\alpha}^{\rho\sigma}C_{\alpha}^{\sigma\rho}\biggr)\tilde{I}+\sum_{\omega\not=\rho}\biggl(\sum_{\sigma\not=\rho,\omega}C_{\alpha}^{\rho\sigma}C_{\alpha}^{\sigma\omega}\biggr)\bigl[C_{\mu}\mu_{\rho\omega}+\ldots\bigr]\,;
\label{example_matrices}
\end{align}
comparing with (\ref{OPE_sigma_aa}) we obtain the matrix relation
\EQ
C_{\alpha}^2=q_1I+q_2C_{\alpha}\,.
\label{C_prod_1}
\EN
Analogously, from (\ref{OPE_sigma_ab}) we derive
\EQ
C_{\alpha}C_{\beta}=-I-(C_{\alpha}+C_{\beta})\,,\hspace{1cm}\alpha\not=\beta\,,
\label{C_prod_2}
\EN
which implies in particular $[C_{\alpha},C_{\beta}]=0$. Hence, the set of $q$ hermitian matrices $C_{\alpha}$ can be simultaneously diagonalized by a unitary transformation $U$ and put in the form
\EQ
\left.C_{\alpha}^{\rho\sigma}\right|_{\text{diag}}=(q\delta_{\alpha\rho}-1)\delta_{\rho\sigma}\,,
\label{C_diag}
\EN 
which follows from the observation that the $C_{\alpha}$'s are traceless, sum to zero and, due to (\ref{C_prod_1}), have $-1$ and $q_1$ as only eigenvalues. It is also simple to check that
\begin{align}
&\langle\mu_{\alpha}\mu_{\alpha}\rangle=\frac{1}{q}\text{Tr}\,C_{\alpha}^2\text{
  }\langle\mu_{\alpha\beta}\mu_{\beta\alpha}\rangle=q_1\langle\mu_{\alpha\beta}\mu_{\beta\alpha}\rangle\,,\\
&\langle\mu_{\alpha}\mu_{\alpha}\mu_{\alpha}\rangle=\frac{1}{q}\text{Tr}\,C_{\alpha}^3\text{
  }\langle\mu_{\alpha\beta}\mu_{\beta\gamma}\mu_{\gamma\alpha}\rangle=q_2q_1\langle\mu_{\alpha\beta}\mu_{\beta\gamma}\mu_{\gamma\alpha}\rangle\,,
\end{align}
in  agreement with (\ref{duality_2pts}), (\ref{duality_3pts}). Starting
from the diagonal form (\ref{C_diag}) and the existence of the unitary
matrix $U$ it is possible to show (see Appendix~C) that
\EQ
C_{\alpha}^{\rho\sigma}=\text{e}^{i(\varphi_{\alpha\sigma}-\varphi_{\alpha\rho})}-\delta_{\rho\sigma},
\label{C_nondiag}
\EN 
where the phases $\varphi_{\alpha\rho}$ must satisfy the equation\footnote{The properties of matrices $C_\alpha$ we obtain in this section do not refer to any specific value of $q$. Of course they are satisfied by the matrices that we already determined in the previous section for $q=2,3,4$.}
\EQ
\frac{1}{q}\sum_{\rho=1}^q\text{e}^{i(\varphi_{\alpha\rho}-\varphi_{\beta\rho})}=\delta_{\alpha\beta}\,.
\label{phases}
\EN
We can now notice that (\ref{mapping}), (\ref{Gdual}) and (\ref{duality_mapping}) imply
\EQ
q_1D_{\alpha_1\alpha_2\alpha_3\alpha_4}^{\alpha\beta\alpha\beta}=\frac{1}{q}\sum_{\rho,\sigma}C_{\alpha_1}^{\rho\sigma}C_{\alpha_2}^{\sigma\rho}C_{\alpha_3}^{\rho\sigma}C_{\alpha_4}^{\sigma\rho}\,;
\label{Dabab_eq}
\EN   
using $|C_{\alpha}^{\rho\sigma}|=1-\delta_{\rho\sigma}$ and hermiticity $C_{\alpha}^{\sigma\rho}=\left(C_{\alpha}^{\rho\sigma}\right)^{*}$ we obtain $D_{\alpha\alpha\alpha\alpha}^{\alpha\beta\alpha\beta}=D_{\alpha\alpha\beta\beta}^{\alpha\beta\alpha\beta}=D_{\alpha\beta\beta\alpha}^{\alpha\beta\alpha\beta}=1$,
  independently of the phases $\varphi_{\alpha\beta}$. With this information, (\ref{4pts_OPE1}) and (\ref{4pts_OPE2}) determine the remaining coefficients, giving
\begin{align}
&G_{\alpha\alpha\alpha\alpha}=q_1q_2q_3\,\tilde{G}_{\alpha\beta\gamma\delta}+q_
1q_2\,\left(G_{\alpha\beta\gamma\beta}+\tilde{G}_{\alpha\beta\alpha\gamma}\right)+q_1\,\tilde{G}_{\alpha\beta\alpha\beta},\label{duality_aaaa}\\
&G_{\alpha\alpha\beta\beta}=-q_2q_3\,\tilde{G}_{\alpha\beta\gamma\delta}-q_2\,\tilde{G}_{\alpha\beta\gamma\beta}+q_1q_2\,\tilde{G}_{\alpha\beta\alpha\gamma}+q_1\,\tilde{G}_{\alpha\beta\alpha\beta},\label{duality_aabb}\\
&G_{\alpha\beta\beta\alpha}=-q_2q_3\,\tilde{G}_{\alpha\beta\gamma\delta}+q_1q_2\,\tilde{G}_{\alpha\beta\gamma\beta}-q_2\,\tilde{G}_{\alpha\beta\alpha\gamma}+q_1\,\tilde{G}_{\alpha\beta\alpha\beta},\label{duality_abba}
\end{align}
with different indices denoting different colors.

The problem with the remaining correlator $G_{\alpha\beta\alpha\beta}$ is that (\ref{Dabab_eq}) does not help, because the phases do not cancel, and we are left with the three equations coming from the OPE for four unknowns.

\section{The boundary case}
Since the OPE's for the kink and spin fields reflect local properties of the field theory, the duality relations obtained in the previous section hold true also in the case in which the points $x_1,\ldots,x_n$ in (\ref{duality_mapping}), instead of being located on the infinite plane, lie inside a simply connected domain $L\subset\mathbb{R}^2$. Actually, the duality relations continue to hold also in the case the points $x_	1,\ldots,x_n$ are located on the boundary of $L$, simply because the OPE's (\ref{kink_OPE}), (\ref{OPE_sigma_ab}) and (\ref{OPE_sigma_aa}), whose structure is completely determined by the symmetry, can be used also for points on the boundary, provided bulk OPE coefficients and scaling dimensions are replaced by boundary OPE coefficients and scaling dimensions. It is not difficult to see, however, that having the points on the boundary rather than in the bulk may reduce the number of linearly independent correlation functions. The first interesting case, that we now discuss, arises for $n=4$.

\begin{figure}[t]
\begin{center}
\includegraphics[height=4.5cm]{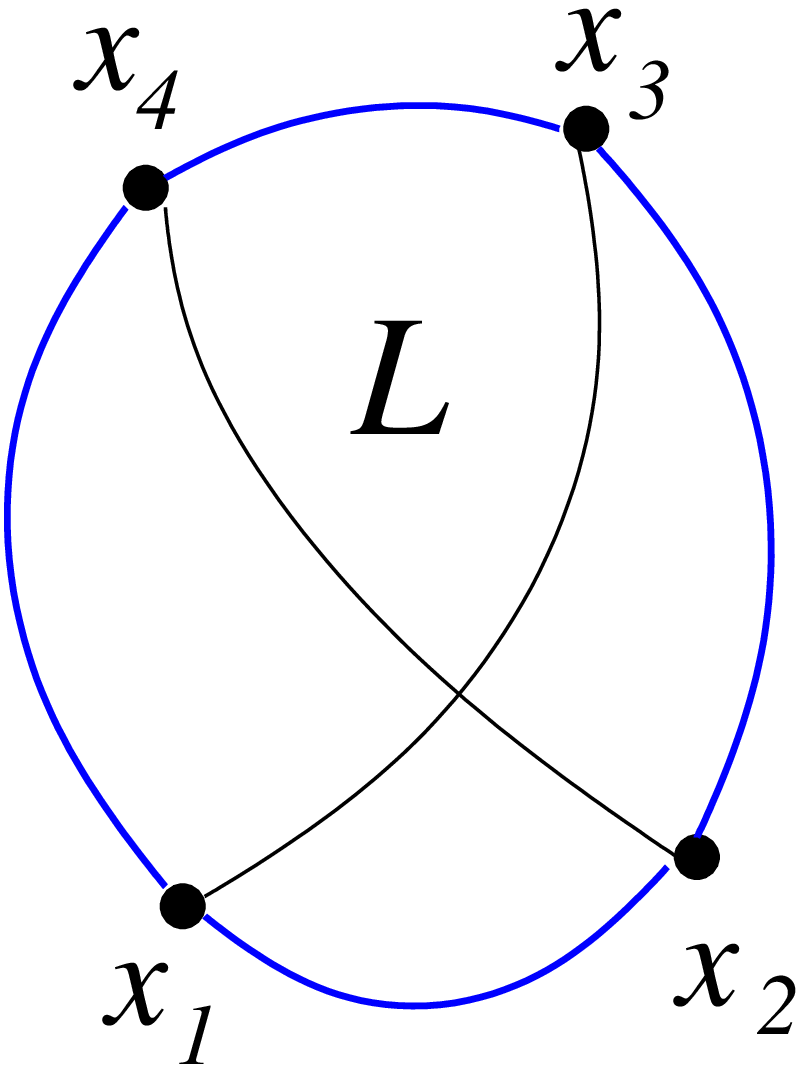}\hspace{1.5cm}
\includegraphics[height=5.1cm]{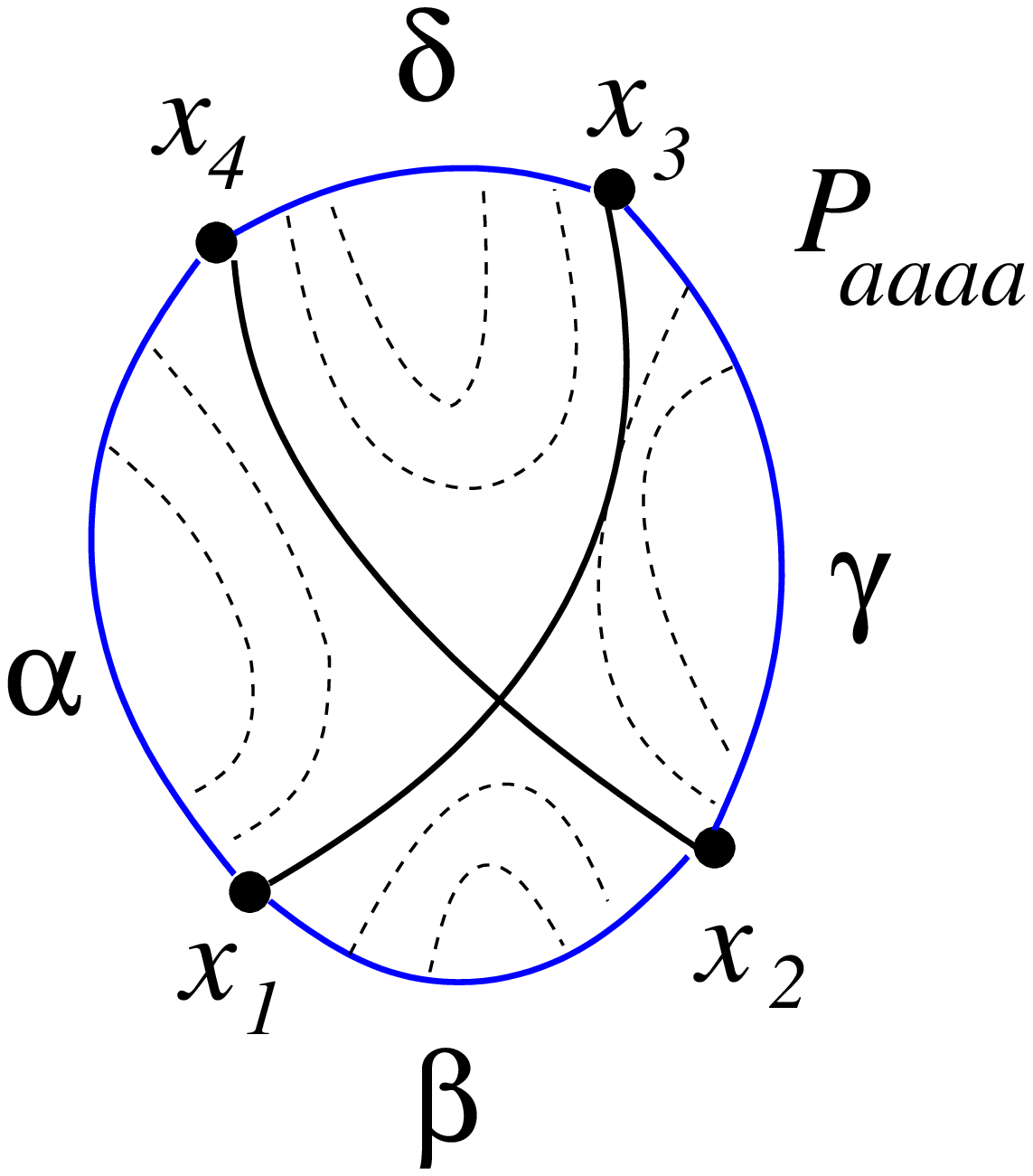}
\caption{Four-point boundary correlations on a simply connected domain $L$. \textbf{Left: } Clusters connecting $x_1$ to $x_3$ and $x_2$ to $x_4$ necessarily cross and cannot be distinct. \textbf{Right: }Dashed lines represent allowed clusters in the FK representation on the dual lattice $\mathcal{L}^{*}$; continuous lines correspond to clusters on $\mathcal{L}$ connecting all the four points. In the continuum limit $\mathcal{L}$, $\mathcal{L}^{*}$ and $L$ coincide.}
\label{fig_boundary}
\end{center}
\end{figure}

Let us order the points $x_1,\ldots,x_4$ on the boundary as shown in Fig.~\ref{fig_boundary}. Since in this boundary case a cluster containing $x_1$ and $x_3$ must necessarily cross a cluster containing $x_2$ and $x_4$, the probability $P_{abab}(x_1,x_2,x_3,x_4)$ that these two pairs of points belong to two different clusters necessarily vanishes. This topological  constraint reduces to three the number of
linearly independent boundary spin correlators (we denote them with a superscript $B$); indeed, inverting (\ref{prob_aaaa})--(\ref{prob_abab}) and setting $P_{abab}=0$ gives\footnote{Different greek indices in eqs. (\ref{sumrule_Potts_spin}--\ref{Pabba_kink}) denote different colors.}
\EQ
q_1(q^2-3q+1)G^B_{\alpha\beta\alpha\beta}-(2q-3)G^B_{\alpha\alpha\alpha\alpha}+q_1(G^B_{\alpha\alpha\beta\beta}+G^B_{\alpha\beta\beta\alpha})=0\,.
\label{sumrule_Potts_spin}
\EN     
Using the duality relations  (\ref{duality_aaaa}), (\ref{duality_aabb}), (\ref{duality_abba}) and the OPE equations (\ref{4pts_OPE1}) for $G_{\alpha\beta\alpha\beta}$, (\ref{sumrule_Potts_spin})  can be rewritten as 
\EQ
\bigl[1+q_1(q^2-3q+1)D_{\alpha\beta\alpha\beta}^{\alpha\beta\alpha\beta}\bigr]\bigl(\tilde{G}^B_{\alpha\beta\alpha\beta}+\tilde{G}^B_{\alpha\beta\gamma\delta}-\tilde{G}^B_{\alpha\beta\alpha\gamma}-\tilde{G}^B_{\alpha\beta\gamma\beta}\bigr)=0\,,
\label{sum_rule_kink}
\EN 
or, in view of the reduction in the number of independent correlators , as the linear relation\footnote{Duality for boundary correlators of the $q$-state Potts model on the lattice was studied in \cite{Wu_duality,Jacobsen,WH}, where dual partition functions with domain wall boundary conditions correspond to our kink field boundary correlators $\tilde{G}^B_{\alpha_1\alpha_2,\ldots}$; with this identification, the relation (\ref{LSZ}) is contained in \cite{WH}. Potts partition functions on a non-simply connected domain have been studied in \cite{King}. An early investigation of Potts correlation functions is in \cite{Wu_Graphs}.}
\EQ
\tilde{G}^B_{\alpha\beta\alpha\beta}+\tilde{G}^B_{\alpha\beta\gamma\delta}=\tilde{G}^B_{\alpha\beta\alpha\gamma}+\tilde{G}^B_{\alpha\beta\gamma\beta}.
\label{LSZ}
\EN
Using this equation to eliminate $\tilde{G}^B_{\alpha\beta\alpha\beta}$ in (\ref{duality_aaaa}),
(\ref{duality_aabb}) and (\ref{duality_abba}), the duality for boundary correlators is fully determined as
\begin{align}
&G^B_{\alpha\alpha\alpha\alpha}=q_1(q_2q_3-1)\tilde{G}^B_{\alpha\beta\gamma\delta}+q_1^2(\tilde{G}^B_{\alpha\beta\gamma\beta}+\tilde{G}^B_{\alpha\beta\alpha\gamma}),\label{bd_aaaa}\\
&G^{B}_{\alpha\alpha\beta\beta}=(2q_2-q_1^2)\tilde{G}^B_{\alpha\beta\gamma\delta}+\tilde{G}^B_{\alpha\beta\gamma\beta}+q_1^2\tilde{G}^B_{\alpha\beta\alpha\gamma},\label{bd_aabb}\\
&G^{B}_{\alpha\beta\beta\alpha}=(2q_2-q_1^2)\tilde{G}^B_{\alpha\beta\gamma\delta}+q_1^2\tilde{G}^B_{\alpha\beta\gamma\beta}+\tilde{G}^B_{\alpha\beta\alpha\gamma}\label{bd_abba}.
\end{align}
Substituting in (\ref{prob_aaaa})--(\ref{prob_abab}) with $P_{abab}=0$ one also obtains for the boundary connectivities the simple relations
\begin{align}
&P^B_{aaaa}=\tilde{G}^B_{\alpha\beta\gamma\delta},\label{Paaaa_kink}\\
&P^B_{aabb}=\tilde{G}^B_{\alpha\beta\alpha\gamma}-\tilde{G}^B_{\alpha\beta\gamma\delta},\label{Paabb_kink}\\
&P^B_{abba}=\tilde{G}^B_{\alpha\beta\gamma\beta}-\tilde{G}^B_{\alpha\beta\gamma\delta}\label{Pabba_kink}.
\end{align}
Equation (\ref{Paaaa_kink}) can be interpreted as follows in the language of lattice duality. If we interpret the insertion of a kink field on the boundary as creating a domain wall along the boundary of the dual lattice $\mathcal{L}^{*}$, the correlator $\tilde{G}^B_{\alpha\beta\gamma\delta}$ corresponds to a partition of the boundary into four regions with different colors and will receive contributions only from graphs on $\mathcal{L}^{*}$ without FK clusters connecting different regions. Equation (\ref{Paaaa_kink}) then means that these graphs are in one-to-one correspondence with the graphs on $\mathcal{L}$ in which the four boundary points all belong to the same FK cluster (see Fig.~\ref{fig_boundary})). Similar reasonings can be used for (\ref{Paabb_kink}) and (\ref{Pabba_kink}).

\begin{figure}[t]
\begin{center}
\includegraphics[height=5cm]{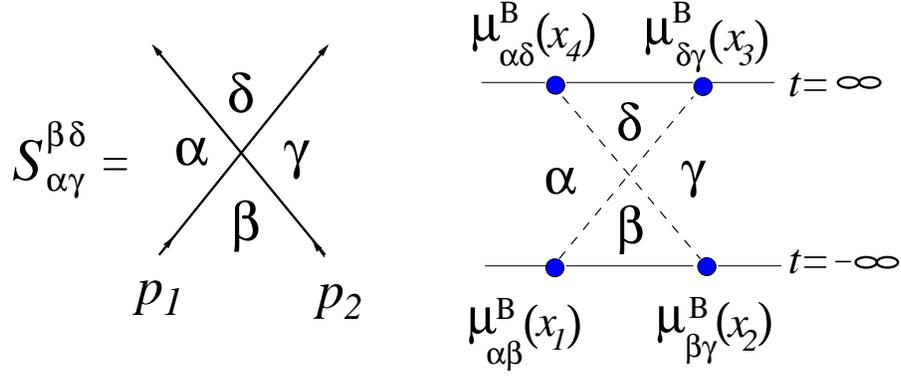}
\caption{The elastic kink-kink scattering amplitudes $S_{\alpha\gamma}^{\beta\delta}(s)$ (left) are related by the LSZ formalism to the four-point correlation functions $\tilde{G}^B_{\alpha\beta\gamma\delta}$ of the asymptotic fields $\mu_{\alpha\beta}^B(x)$ (right).}
\label{fig_LSZ}
\end{center}
\end{figure}

We conclude this section observing that (\ref{LSZ}) leads to a linear relation among the elastic kink-kink scattering amplitudes in the ($1+1$)-dimensional field theory associated by Wick rotation to the Euclidean theory on the plane. In this case, indeed, the boundary fields $\mu_{\alpha\beta}^B(x)$ entering (\ref{LSZ}) can be interpreted as the asymptotic fields which create a kink excitation $K_{\alpha\beta}$ at time $t\to\pm\infty$ (Fig.~\ref{fig_LSZ}). The kink-kink elastic amplitudes 
\EQ
S_{\alpha\gamma}^{\beta\delta}(s)= {}_{\text{out}}\langle
  K_{\alpha\delta}(p_4)K_{\delta\gamma}(p_3)|K_{\alpha\beta}(p_1)K_{\beta\gamma}(p_2)\rangle_{\text{in}}\,,
\EN
where the square of the center of mass energy $s=(p_1+p_2)_\mu^2$ is the only relativistic invariant for this ($1+1$)-dimensional process, can be written within the LSZ formalism (see e.g. \cite{IZ}) as
\begin{equation}
\begin{split}
S_{\alpha\gamma}^{\beta\delta}(s)= &\lim_{t_1,t_2\to-\infty}\,\,\,\lim_{t_3,t_4\to+\infty}\int\text{d}x_1\dots\int\text{d}x_4\text{
  e}^{ip_4\cdot x_4}\text{e}^{ip_3\cdot x_3}\text{e}^{-ip_1\cdot x_1}\text{e}^{-ip_2\cdot x_2}\\
&\overleftrightarrow{\partial_{t_4}}\text{
}\overleftrightarrow{\partial_{t_3}}\text{
}\overleftrightarrow{\partial_{t_2}}\text{
}\overleftrightarrow{\partial_{t_1}}\text{ }\langle\mu^B_{\alpha\beta}(x_1,
t_1)\mu^B_{\beta\gamma}(x_2, t_2)\mu^B_{\gamma\delta}(x_3, t_3)\mu^B_{\delta\alpha}(x_4,t_4)\rangle\,,
\end{split}
\label{LSZ_explicit}
\end{equation}
where the integrals are taken along the one-dimensional space coordinate and $\overleftrightarrow{\partial_t}$ is defined by $A\overleftrightarrow{\partial_t}B=A({\partial_t}B)-({\partial_t}A)B$. Equation (\ref{LSZ}) then leads to
\EQ
S_{\alpha\gamma}^{\beta\delta}(s)+S_{\alpha\alpha}^{\beta\beta}(s)=S_{\alpha\gamma}^{\beta\beta}(s)+S_{\alpha\alpha}^{\beta\gamma}(s)\,,
\label{rule}
\EN
where different indices denote different colors.
This relation was already found in \cite{CZ} as a byproduct of the Yang-Baxter equations, i.e. of the integrability of the scaling limit of the $q$-state Potts model. Here integrability does not appear to play a role, and then (\ref{rule}) should hold also for non-minimal, non-integrable realizations of the symmetry, if they exist. Ultimately, the vanishing of the probability that the trajectories of two particles do not cross in the ($1+1$)-dimensional space-time relates amplitudes which would be independent on the basis of color symmetry alone.

\section{Conclusion}
In this paper we considered the $q$-color Potts field theory characterized by invariance under color permutations and having the spin fields $\sigma_\alpha(x)$ as fundamental fields. We discussed how the theory can be given a meaning for real positive values of $q\leq q_c$ for the purpose of describing the scaling limit associated to the second order phase transition of the random cluster model, consistently with the expectations coming from the Fortuin-Kasteleyn correspondence. In particular we showed that, although the number of spin fields is badly defined for $q$ non-integer, the number of linearly independent $n$-point spin correlation functions is a $q$-independent integer $F_n$ coinciding with the number of partitions of $n$ elements into subsets each containing more than one element. These $n$-point spin correlators, which in turn determine the connectivities in the random cluster model, contain $n_c\leq n$ different color indices. Hence, they enjoy an ordinary definition for integer values of $q\geq n_c$, but formal use of the symmetry unambiguously prescribes their analytic continuation to real values of $q$. The mechanism allowing the analytic continuation is exemplified by the form of OPE that we obtained for the spin fields (equations (\ref{OPE_sigma_ab}), (\ref{OPE_sigma_aa})): given two colors, the symmetry only discriminates whether they are equal or different, and $q$ enters as a parameter which can take real values.

We also discussed how the purely magnetic properties of the Potts model, which are well defined for integer values of $q$, are described by the sector of the theory in which the only spin variables $\sigma_\alpha$ playing a role are those with $\alpha=1,\ldots,q$. The number $M_n(q)$ of linearly independent $n$-point correlators of such variables, which is smaller than $F_n$ for $n$ large enough, has been determined for $q=3,4$. 

For the two-dimensional case, we studied the duality relations between the spin correlators in the symmetric phase and the kink field correlators in the spontaneously broken phase, both for bulk and boundary correlators, exploiting the OPE as only input. For the case of four-point correlators, the first with $F_n>1$, our method gives the duality relations for three of the four independent bulk correlators, while it completely fixes the duality for independent boundary correlators, whose number is reduced to three for topological reasons. The ultimate reason for our inability to fix completely the bulk duality is that the OPE (\ref{kink_OPE}) for kink fields determines the fusion of adjacent fields only: while this is the only possibility on the boundary, in the bulk also the fusion of non-adjacent fields (i.e. $x_1\to x_3$ in Fig.~\ref{pinching}) is allowed. Clearly this point requires further investigation.

Finally, we showed how the constraints that topology imposes on boundary correlators leads, through the LSZ formalism, to relations among kink scattering amplitudes in $(1+1)$-dimensional space-time. These relations, already observed within a framework based on integrability, appear in our derivation as a generic feature of two-dimensional field theories with spontaneously broken $S_q$ symmetry.

\vspace{1cm} 
\noindent
\textbf{Acknowledgments.} J.V. thanks A. De Luca and F. Mancarella for discussions. Work supported in part by the MIUR Grant 2007JHLPEZ.

\section*{Appendix A} 
Given a set $S$ of $n$ elements the number of partitions of the elements of
$S$ is the Bell number $B_n$ (see Table~\ref{tab_Bell} for $n\leq 10$). The most straightforward way to compute the Bell numbers is through the recursive relation
\EQ
B_n=\sum_{k=0}^{n-1}\binom{n-1}{k}B_k,\quad B_0\equiv1,
\label{bell_def}
\EN 
which is easily proved observing that we can fix one element of the set $S$ and consider the partitions in which this element appears with $k$ other elements, $k=0,\ldots,n-1$. The number of such partitions will be
\EQ
\binom{n-1}{k}B_{n-1-k}\,;
\EN
then summing over $k$ and using
$\binom{n-1}{k}=\binom{n-1}{n-1-k}$ we obtain (\ref{bell_def}).

We similarly define the numbers $F_n$ as
\EQ
B_{n}=\sum_{k=0}^n\binom{n}{k}F_k,\quad F_0\equiv 1.
\label{f_def_2}
\EN 
$F_n$ is the number of partitions of $S$ whose blocks contain at least two
elements. The proof is again elementary (see e.g. \cite{PS}). We divide all the
$B_n$ partitions of $S$ into those containing exactly $k=0,1,\ldots,n$ isolated points, and
then we take partitions of the remaining  $n-k$ points in such way that no point is
isolated. It is clear that we end up with (\ref{f_def_2}).

Recalling the combinatorial identity $\binom{n+1}{k}=\binom{n}{k}+\binom{n}{k-1}$,  expression (\ref{f_def_2}) implies
\EQ
B_{n+1}-B_{n}=\sum_{k=0}^{n}\binom{n}{k}F_{k+1}\,.
\label{proof_equivalence}
\EN
On the other hand, (\ref{bell_def}) gives $B_{n+1}=\sum_{k=0}^{n}\binom{n}{k}B_k$, and
using (\ref{f_def_2}) we derive
\EQ
\sum_{k=0}^n\binom{n}{k}(F_{k+1}+F_k-B_k)=0\quad\forall n\geq 0\,.
\EN 
By induction we finally obtain\footnote{Alternatively, one can recover (\ref{Fdef}) introducing the exponential generating functions $\mathcal{B}(x)=\sum_{n}\frac{B_n}{n!}x^n=\text{e}^{\text{e}^x-1}$ and $\mathcal{F}(x)=\sum_{n}\frac{F_n}{n!}x^n=\text{e}^{\text{e}^x-1-x}$; the result then follows from $\mathcal{B}(x)=\mathcal{F}(x)+\mathcal{F}'(x)$. We thank the referee for this observation.}
\EQ
B_n=F_n+F_{n+1}\,.
\label{Fdef}
\EN

The number of $k$-partitions (partitions into $k$ non-empty subsets) of a set of $n$ elements is the Stirling number $S(n,k)$. It satisfies $B_n=\sum_{k=1}^n S(n,k)$  from its definition, as well as the recursive equation
\begin{equation}
S(n,k)=k\,S(n-1,k)+S(n-1,k-1),\quad\hspace{.5cm} n\geq k,\hspace{.5cm}k\geq 1\,,
\end{equation}
from the fact  that we can obtain a $k$-partition of the set $\{x_1,\dots,x_n\}$ adding $x_n$ to one of the $k$ blocks of a $k$-partition of the elements  $\{x_1,\dots, x_{n-1}\}$, or joining $x_n$ as a single block  to a  $(k-1)$-partition of $\{x_1,\dots, x_{n-1}\}$ . The exponential generating function $\mathcal{S}_k(x)=\sum_{n\geq k} S(n,k)\frac{x^n}{n!}$ satisfies $\mathcal{S}_k'(x)=k\mathcal{S}_k(x)+\mathcal{S}_{k-1}(x)$, and is given by $\mathcal{S}_k(x)=\frac{1}{k!}(\text{e}^{x}-1)^k$. We also introduce the generalized Stirling number $\tilde{S}(n,k)$ as the number of $k$-partitions of a set of $n$ elements whose blocks contain at least two elements (non-singleton $k$-partition); the relation
\EQ
F_n=\sum_{k=1}^{n-1} \tilde{S}(n,k)
\label{F-S}
\EN
then expresses the decomposition of the total number of independent $n$-point spin correlation functions (which we take without isolated indices) into subsets with indices of $k$ different colors. Non-singleton $k$-partitions of the set $\{x_1,\dots, x_n\}$  are obtained adding $x_n$ to one of the $k$ blocks of a non-singleton $k$-partition of  $\{x_1,\dots, x_{n-1}\}$, or by joining the block $\{x_n, x_j\}$, for $j=1,\dots, n-1$, to a non-singleton $(k-1)$-partition of $\{x_1,\dots, x_{j-1}, x_{j+1},\dots,x_{n-1}\}$. We have then
\begin{equation}
\tilde{S}(n,k)=k\tilde{S}(n-1,k)+(n-1)\tilde{S}(n-2,k-1),\quad n\geq k,\text{ }k\geq 1.
\end{equation}
The exponential generating function is $\tilde{\mathcal{S}}_k(x)=\frac{1}{k!}(\text{e}^{x}-1-x)^k$ and solves  $\tilde{\mathcal{S}}'_k(x)=k\tilde{\mathcal{S}}_k(x)+x\tilde{\mathcal{S}}_{k-1}(x)$. The first few $\tilde{S}(n,k)$ are collected in  Table \ref{tab_Stirling}.  
\begin{table}[t]
\begin{center}
\begin{tabular}{|c||c|c|c|c|c|c|c|c|c|c|}
\hline
$n$ & 1 & 2 & 3 & 4 & 5 & 6 & 7 & 8 & 9 & 10  \\
\hline
$F_n$ & 0 & 1 & 1 & 4 & 11 & 41 & 162 & 715 & 3425 & 17722 \\
$\tilde{S}(n,1)$ & 0 & 1 & 1 & 1 & 1 & 1 & 1 & 1 & 1 & 1    \\
$\tilde{S}(n,2)$ & 0 & 0 & 0 & 3 & 10 & 25 & 56 & 119 & 246 & 501    \\
$\tilde{S}(n,3)$ & 0 & 0 & 0 & 0 & 0 & 15 & 105 & 490 & 1918 & 6825 \\
$\tilde{S}(n,4)$ & 0 & 0 & 0 & 0 & 0 & 0 & 0 & 105 & 1260 & 9450 \\
$\tilde{S}(n,5)$  & 0 & 0 & 0 & 0 & 0 & 0 & 0 & 0 & 0 & 945 \\
\hline
\end{tabular}
\caption{The $F_n$ independent $n$-point spin correlators without isolated indices decompose into subsets containing $\tilde{S}(n,k)$ correlators with indices of $k$ different colors.}
\label{tab_Stirling}
\end{center}
\end{table}

\section*{Appendix B}
In this appendix we determine the number $M_n(q)$ of $S_q$-inequivalent, linearly independent $n$-point spin correlators (\ref{G}) which determine magnetic correlations in the Potts model at $q=2,3,4$. We exploit the fact, discussed in Section~3, that at $q=2,3,4$ the symmetric group factorizes as
\EQ
S_q=K_q\rtimes S_{q-1}\,,
\EN
with $K_4=D_2$, $K_3=\mathbb Z_3$, $K_2=\mathbb{Z}_2$, and that the Potts model is described by  $q-1$ independent spin variables $t_1,\dots, t_{q-1}$ charged under the abelian group $K_q$. Non-vanishing  correlation functions are neutral under $K_q$ and invariant under permutations of the $q-1$ operators $t_i$.

At $q=2$, the only independent variable is  $t_1$ with charge +1 under $\mathbb
Z_2$. The neutrality condition for the $n$-point correlation functions is
\EQ
n\equiv 0\mod 2,
\label{const_q2}
\EN
giving $M_{2k}(2)=1$, $M_{2k+1}(2)=0$.

At $q=3$, the independent spin variables are $t_1$ and
$t_2$ with $\mathbb Z_3$ charge +1 and $-1$, respectively. Given a $n$-point correlation
function containing $n_1$ variables $t_1$ and $n_2$ variables $t_2$ we require
\begin{align}
& n_1+n_2=n,\\
& n_1-n_2\equiv 0\mod 3,
\label{const_q3}
\end{align}
or equivalently $n_1+n\equiv 0\mod 3$, with $n_1=0,\dots,n$. Assigned the
couple of integers $\{n_1, n_2\}=\{n_1, n-n_1\}$ satisfying the constraint (\ref{const_q3}),
  the total number of distinct correlation functions we can construct is the
  binomial coefficient $\binom{n}{n_1}$. The charge conjugation operation, however, exchanges $n_1$ with $n_2$, and correlation functions obtained by
  $n_1\rightarrow n-n_1$ are equal; notice indeed that
  $\binom{n}{n_1}=\binom{n}{n-n_1}$. It follows\footnote{If $n$ is even and $n_1=n/2$ the factor 1/2 in (\ref{Mn_3})  avoids the double counting of the correlation functions obtained exchanging in block the positions of the $n_1$ operators $t_1$ with the $n_2=n_1$ operators $t_2$.}
\EQ
M_n(3)=\frac{1}{2}\sum_{\substack{n=0\\n_1+n\equiv 0\\ \mod
    3}}^n\binom{n}{n_1}.
\label{Mn_3}
\EN  
The elements of the sequence (\ref{Mn_3}) (see the first few of them in Table~\ref{tab_Bell}) coincide with the Jacobsthal numbers \cite{OEIS_1} and satisfy the
recursive relation
\EQ
M_{n+1}(3)=M_{n}(3)+2M_{n-1}(3)\,,
\label{rec_Mn_3}
\EN
with $M_1(3)=0$, $M_2(3)=1$. We will not prove (\ref{rec_Mn_3}) directly but we will justify it through the following observation. Consider an
hypertetrahedron with $q$ vertices labeled by the numbers $1,\dots,q$. The number $y^{(n)}$ of closed $n$-step paths
starting from a given vertex, say $1$, of the hypertetrahedron satisfies the
recursive relation (see Appendix C)
\EQ
y^{(n)}=(q-2)y^{(n-1)}+(q-1)y^{(n-2)}\,,
\label{rec_y} 
\EN
with $y^{(1)}=0$, $y^{(2)}=q-1$. The closed $n$-step paths $\gamma^{(n)}$ in (\ref{rec_y}) are considered distinct even when they differ
by a  permutation $\pi\in S_{q-1}$ of the $q-1$ vertices $2,\dots,q$. At $q=3$ 
there is only one possible permutation $\pi$, and it  exchanges the vertices $2$
and $3$.  The application of $\pi$ to a path $\gamma^{(n)}$ generates the path reflected along the
symmetry axis containing the vertex $1$ of an equilateral triangle. The
number of closed paths inequivalent under permutations at $q=3$ is then just half of
the total number of closed paths $y^{(n)}$, and in particular satisfies
(\ref{rec_y}) with $q=3$. Finally notice that closed $n$-step paths $\gamma^{(n)}$ 
inequivalent under permutations are in one to one correspondence with
independent $n$-point kink fields
correlation functions\footnote{Any path $\gamma^{(n)}$ can be represented as the
  sequence of $n+1$ vertices $\{1,v_2,\dots,v_{n},1\}$ with $v_{i}\not=v_{i+1}$
  and $v_{i}=1\dots q$. The associated kink fields correlation function is
  $\langle\mu_{1v_2}(x_1)\dots\mu_{v_n1}(x_n)\rangle$. Alternatively
$\gamma^{(n)}$ can be thought as a particular coloration of $n$ points on a
  circle. See again Appendix C.} (\ref{Gtilde}), 
whose number is  $M_n(q)$. Consistency requires that $M_n(3)$ satisfies the
recursive equation (\ref{rec_y}) with $q=3$, which indeed coincides with
(\ref{rec_Mn_3}).

For $q=4$ we  consider correlation functions of the three variables $t_1, t_2$
and $t_3$ with $D_2=\mathbb Z_2\times\mathbb Z_2$ charges $(1,0)$, $(0,1)$ and
$(1,1)$, respectively. A non-vanishing $n$-point correlation function with $n_1$
variables $t_1$, $n_2$ variables $t_2$ and $n_3$ variables $t_3$ satisfies
\begin{align}
&n_1+n_2+n_3=n, \label{const_4a}\\
&(n_1+n_3, n_2+n_3)\equiv (0,0)\mod 2\,,\label{const_4b}
\end{align}           
or, more symmetrically, $n_i\equiv n\mod 2$, $i=1,2,3$. The number of
distinct $n$-point correlation functions associated to the solution $\{n_1, n_2,
n_3\}$ of (\ref{const_4a}) and (\ref{const_4b}) is
$\frac{n!}{n_1!n_2!n_3!}$. Correlation functions obtained by permutations of the
integers $n_i$ are identified and we must therefore choose a definite order
for them, for example $n_1\leq n_2\leq n_3$; $n_i=0,\ldots,n$. When two positive integers  $n_1$
and $n_2$ coincide, the two correlation functions obtained by exchanging in block the
positions of the $n_1$ operators $t_1$ with the $n_2$ operators $t_2$ are also
equal and  counted twice among the $\frac{n!}{n_1!n_2!n_3!}$
correlation functions. Similarly, if $n_1=n_2=n_3$ permutational symmetry
does not distinguish among  the $3!$ correlation functions obtained
exchanging in block
the positions of the $n_i$ variables $t_i$ for $i=1,2,3$. The final
result is then 
\EQ
M_n(4)=\sum_{\substack{n_1+n_2+n_3=n\\n_i\equiv n\mod
    2}}\frac{n!}{n_1!n_2!n_3!}\,\frac{1}{n_e(n_1,n_2,n_3)!}\,,
\label{Mn_4}
\EN  
where the $n_i$ are ordered, $n_i\leq n_{i+1}$, and $n_e(n_1,n_2,n_3)$ is the number of
non-zero equal integers in the tern $\{n_1, n_2, n_3\}$. The integer sequence
(\ref{Mn_4}) (see Table~\ref{tab_Bell}) is also known \cite{OEIS_2}, and is solution of the recursive equation
\EQ
M_{n+1}(4)=2M_{n}(4)+3M_{n-1}(4)-1\,,
\label{rec_Mn_4}
\EN
with $M_{1}(4)=0$, $M_2(4)=1$. Again we will not prove (\ref{rec_Mn_4}) directly, but will explain
why  a proper counting of closed paths $\gamma^{(n)}$ inequivalent under permutations
of the vertices on a tetrahedron is obtained subtracting 1 to the r.h.s. of
(\ref{rec_y}) with $q=4$. It is convenient to represent $\gamma^{(n)}$
 as the sequence $\gamma^{(n)}=\{1,v_2,\dots,v_n,1 \}$ with $v_i=1,\dots, 4$ and
 $v_{i}\not=v_{i+1}$, which is also a particular coloration of $n$ points on a
 circle. The way an $(n+1)$-step closed path $\gamma^{(n+1)}=\{1,v_2,\dots, v_{n}, v_{n+1}, 1\}$ is constructed by adding a new point $v_{n+1}$ and taking care of permutational symmetry is the following\footnote{Notice that there are two cases corresponding to $v_n=1$ or $v_n\not=1$. In the first case the new point $v_{n+1}$ is added to some closed $(n-1)$-step  path $\gamma^{(n-1)}$; in the second case the new point is added to an $(n-1)$-step open path. The number of $(n-1)$-step open paths is however equal to the number of closed $n$-step paths $\gamma^{(n)}$}. First we identify the largest $v_i$, $i=1,\dots, n$, and then generate all the paths $\gamma^{(n+1)}$ for which $v_{n+1}\not=v_{n}, 1,$ with $v_{n+1}=2,\dots,\min\{\max\{v_i\}+1,4\}$. The recursion (\ref{rec_y}) fails only if $\max\{v_i\}=2$ and then $v_{n}=1,2$. In this case the closed paths $\gamma^{(n+1)}$ with $v_{n+1}=3,4$ are identified by permutational symmetry and the choice $v_{n+1}=4$ must be discarded; this leads to (\ref{rec_Mn_4}).

We conclude noticing that $n$ distinct vertices are sufficient (and necessary) to enumerate all closed $n$-step paths inequivalent under permutations of the vertices on an hypertetrahedron. In particular  the number $M_n(q)$ of  closed $n$-step paths inequivalent under permutations on an hypertetrahedron with $q$ vertices is constant for any $q\geq n$, and we have already shown in Section 2.3 that, when no restriction is assumed on the number of available vertices so that all closed inequivalent paths are counted, this number coincides with $F_n$:
\EQ
M_n(q)=F_n\hspace{.5cm}\mbox{for}\hspace{.3cm}q\geq n\,.
\label{M-F}
\EN

\section*{Appendix C}
Given the diagonal form (\ref{C_diag}) and the unitary matrix $U$, we have
\begin{align}
C_{\alpha}^{\rho\sigma}&=\sum_{\lambda,\nu}U^{\dagger}_{\rho\lambda}(q\delta_{\alpha\nu}-1)\delta_{\lambda\nu}U_{\nu\sigma}\\
&=qU^{\dagger}_{\rho\alpha}U_{\alpha\sigma}-\delta_{\rho\sigma}.
\label{C_nondiag_pre}
\end{align}
Requiring $C_{\alpha}^{\rho\rho}=0$ for $\rho, \alpha=1\dots q$, gives
\EQ
U_{\alpha\rho}=\frac{1}{\sqrt{q}}\text{e}^{i\varphi_{\alpha\rho}};
\label{unitary_U}
\EN
the equation (\ref{phases}) for the phases is then the self-consistent
condition of unitarity of the matrix $U$. Substituting (\ref{unitary_U}) back into (\ref{C_nondiag_pre}) we obtain (\ref{C_nondiag}). The matrices $C_{\alpha}$ in (\ref{mapping_q2}) and (\ref{mapping_q3}) for $q=2,3$ are reproduced by the solution
\EQ
\varphi_{\alpha\rho}=\pm \frac{2\pi}{q}\alpha\rho
\label{phases_zq}
\EN
of the phase equation (\ref{phases}). For $q=4$, the matrix (\ref{mapping_q4}) corresponds  instead\footnote{Of course (\ref{phases_zq}) solves (\ref{phases}) also for $q=4$. This solution, however, would lead to a matrix $C_{\alpha}$ associated to $K_4=\mathbb Z_4$, inconsistent with our general discussion of section~3 (in particular, $\text{Aut}(\mathbb Z_4)=\mathbb Z_2$).} to the solution
\EQ
\varphi_{\boldsymbol{\alpha}\boldsymbol{\rho}}=\pm\pi(\alpha_1\rho_1+\alpha_2\rho_2),
\EN
with $\boldsymbol{\alpha}=(\alpha_1,\alpha_2)$, $\boldsymbol{\rho}=(\rho_1,
\rho_2)$, $\alpha_i, \rho_i=1, 2$. 

We conclude this appendix
giving a simple geometrical interpretation of the relation (\ref{C_prod_1}). Without loss of generality we can choose the phases so that one of the matrices $C_\alpha$, say $C_q$, is real, i.e. 
\EQ
C_q=\begin{pmatrix}
0& 1 &\cdots & 1 \\
1& 0 &\cdots & 1 \\
\vdots&\vdots &\ddots &\vdots \\
1&1 &\cdots &0 
\end{pmatrix}.
\EN
This is the adjacency matrix of a fully connected graph with $q$ sites, i.e. the projection on the plane of a hypertetrahedron with $q$ vertices. We can then use the formula (\ref{C_prod_1}) to compute powers of $C_q$ and obtain
\EQ
C_q^n=y^{(n)}I+x^{(n)}C_q,
\EN
with the integers $x^{(n)}$ and $y^{(n)}$ satisfying the recursive equations
\begin{align}
&y^{(n+1)}=q_1x^{(n)}, \label{rec1}\\
&x^{(n+1)}=q_2x^{(n)}+y^{(n)}.\label{rec2}
\end{align}
It is now simple to realize that $y^{(n)}$ and $x^{(n)}$ are, respectively, the number of closed paths of length $n$ starting from a vertex $v_1$ of the hypertetrahedron and the number of open paths of length $n$ from $v_1$ to $v_k$. Indeed, if we consider such an open path and we add the link $(v_k, v_1)$ we obtain a closed path of length $n+1$; however, by permutational symmetry, all the locations of $v_k\not=v_1$ are equivalent, and we obtain (\ref{rec1}). Suppose instead to remove from the original open path the last link $(v_j, v_k)$; this gives a path of length $n-1$ from $v_1$ to $v_j$ that can be open with multiplicity $q_2$ ($j\not = 1,k$), or closed with multiplicity one ($j=1$), reproducing (\ref{rec2}). The recursions (\ref{rec1}) and (\ref{rec2}) give for $y^{(n)}$ equation (\ref{rec_y}), whose solution is
\EQ
y^{(n)}=\frac{1}{q}\bigl[(-1)^nq_1+q_1^n\bigr],
\EN
which is $q^{-1}$ times the chromatic polynomial\footnote{Given a graph $\mathcal{G}$, the
  chromatic polynomial $\pi_{\mathcal{G}}(q)$ is the number of possible
  colorations of the vertices of $\mathcal{G}$ with $q$ colors. Adjacent
  vertices have different colors and colorations differing by permutations of
  the colors are considered distinct.} $\pi_{C_n}(q)$ for the the cyclic graph
$C_n$, i.e. the graph obtained putting $n$ points on a circle. We conclude that the decompositions over kink fields of the correlator $\langle\mu_\alpha(x_1)\mu_\alpha(x_2)\ldots\mu_\alpha(x_n)\rangle$ can be associated to closed $n$-step paths on a fully connected graph with $q$ vertices, or to colorations of a ring of $n$ points with $q$ colors.

\end{document}